\documentclass[useAMS,usenatbib]{mn2e}
\usepackage{graphicx}
\usepackage{hyperref}
\usepackage{tikz}
\usetikzlibrary{calc,fadings,decorations.pathreplacing}

\newcommand\pgfmathsinandcos[3]{%
  \pgfmathsetmacro#1{sin(#3)}%
  \pgfmathsetmacro#2{cos(#3)}%
}
\newcommand\LongitudePlane[3][current plane]{%
  \pgfmathsinandcos\sinEl\cosEl{#2} 
  \pgfmathsinandcos\sint\cost{#3} 
  \tikzset{#1/.style={cm={\cost,\sint*\sinEl,0,\cosEl,(0,0)}}}
}
\newcommand\LatitudePlane[3][current plane]{%
  \pgfmathsinandcos\sinEl\cosEl{#2} 
  \pgfmathsinandcos\sint\cost{#3} 
  \pgfmathsetmacro\yshift{\cosEl*\sint}
  \tikzset{#1/.style={cm={\cost,0,0,\cost*\sinEl,(0,\yshift)}}} %
}
\newcommand\DrawLongitudeCircle[2][1]{
  \LongitudePlane{\angEl}{#2}
  \tikzset{current plane/.prefix style={scale=#1}}
  \pgfmathsetmacro\angVis{atan(sin(#2)*cos(\angEl)/sin(\angEl))} %
  \draw[current plane] (\angVis:1) arc (\angVis:\angVis+180:1);
  \draw[current plane,dashed] (\angVis-180:1) arc (\angVis-180:\angVis:1);
}

\newcommand\DrawLatitudeCircle[2][1]{
  \LatitudePlane{\angEl}{#2}
  \tikzset{current plane/.prefix style={scale=#1}}
  \pgfmathsetmacro\sinVis{sin(#2)/cos(#2)*sin(\angEl)/cos(\angEl)}
  \pgfmathsetmacro\angVis{asin(min(1,max(\sinVis,-1)))}
  \draw[current plane] (\angVis:1) arc (\angVis:-\angVis-180:1);
  \draw[current plane,dashed] (180-\angVis:1) arc (180-\angVis:\angVis:1);
}

\tikzset{%
  >=latex, 
  inner sep=0pt,%
  outer sep=2pt,%
  mark coordinate/.style={inner sep=0pt,outer sep=0pt,minimum size=3pt,
    fill=black,circle}%
}

\title[FermiFAST]{A Fast Algorithm for Finding Point Sources
in the Fermi Data Stream: FermiFAST}
\author[A. Asvathaman {\em et al.}]{Asha Asvathaman$^{1}$, Conor Omand$^{1}$, Alistair Barton$^{2}$,
  Jeremy S. Heyl$^{1}$\thanks{Email: heyl@phas.ubc.ca; Canada Research Chair} \\
$^{1}$Department of Physics and Astronomy, University of British
  Columbia, 6224 Agricultural Road, Vancouver, BC V6T 1Z1, Canada\\
$^{2}$Department of Physics, McGill University, Montr´eal, QC, H3A 2T8, Canada}

\begin{document}
\date{Accepted ---. Received ---; in original form ---}

\pagerange{\pageref{firstpage}--\pageref{lastpage}} \pubyear{2015}

\maketitle

\label{firstpage}

\begin{abstract}
  We present a new and efficient algorithm for finding point sources
  in the photon event data stream from the Fermi Gamma-Ray Space
  Telescope, FermiFAST. The key advantage of FermiFAST is that it
  constructs a catalogue of potential sources very fast by arranging
  the photon data in a hierarchical data structure. Using this
  structure FermiFAST rapidly finds the photons that could have
  originated from a potential gamma-ray source.  It calcuates a
  likehihood ratio for the contribution of the potential source using
  the angular distribution of the photons within the region of
  interest.  It can find within a few minutes the most significant
  half of the Fermi Third Point Source catalogue (3FGL) with nearly
  80\% purity from the four years of data used to construct the
  catalogue.  If a higher purity sample is desirable, one can achieve
  a sample that includes the most significant third of the Fermi 3FGL
  with only five percent of the sources unassociated with Fermi
  sources.  Outside the galaxy plane, all but eight of the 580
  FermiFAST detections are associated with 3FGL sources.  And of these
  eight, six yield significant detections of greater than five sigma
  when a further binned likelihood analysis is performed.  This
  software allows for rapid exploration of the Fermi data, simulation
  of the source detection to calculate the selection function of
  various sources and the errors in the obtained parameters of the
  sources detected.
  
\end{abstract}


\section{Introduction}

The launch of the Fermi Gamma-Ray Telescope on 11 June 2008 brought
dramatically more sensitive instruments to bear on the study of the
Universe in gamma rays.  The observations of the Fermi LAT (Large Area
Telescope) \citep[e.g.][]{2012ApJS..203....4A} have resulted in a
series of catalogues of point sources both within the Galaxy and
beyond \citep[e.g.][]{2010ApJS..188..405A,2011ApJ...743..171A,
  2012ApJS..199...31N,2013ApJS..208...17A,
  2013ApJS..209...34A,2015ApJS..218...23A}.  Both the identification
of sources and their characterisation rely on the calculation of a
likelihood ratio for the observed data with and without a source and
for different characteristics for the potential source
\citep{1979ApJ...228..939C,1992MNRAS.259..413S,1996ApJ...461..396M}.
These telescopes as well as Cerenkov telescopes on the ground rely on
the conversion of gamma-rays into electron-positron pairs as they
enter the telescope (or the atmosphere).  By tracing the motion of
these pairs, one can reconstruct the momentum of the incoming gamma
ray.  For low gamma-ray energies the reconstruction is poorer than at
higher energies.  Because of this broad point-spread function of
gamma-ray telescopes the contribution of several potential sources
will overlap in a particular region of the sky making the generation
of a catalogue even more difficult and time consuming.

Although both Fermi and EGRET before it used the likelihood ratio to
assess the significance of potential point sources, several
alternatives have been proposed to search for point sources to account
for the photons observed by these instruments.
\citet{2009arXiv0912.3843M} and \citet{2008MNRAS.383.1166C} developed an algorithm
to build a minimal spanning tree from the arrival directions of the
photons over a large portion of the sky.  They then divide the tree
into subclusters by removing branches larger than a particular
threshold.  If a subcluster contains more than a particular number of
photons, it is considered to be a potential point source.
\citet{2013Ap&SS.347..169C} extended the technique to include a noisy
background.
\citet{1997ApJ...483..350D,1997ApJ...483..370D} and \citet{2007AIPC..921..546C}
proposed to use wavelets to find point sources from photon counts for
ROSAT, EGRET and Fermi data.  Althoguh these techniques are
significantly faster than the traditional likelihood ratio, they
require simulated data to assess the significance of the sources
detected because they do not rely of statistics with well understood
distributions given the null hypothesis of no point source at a
particular location.  In general the estimated significances are
proportional to those found in the full likelihood analysis.

In this paper we propose a technique to calculate likelihood ratios
much more quickly that the usual techniques by storing the photon data
stream in a data structure optimized for finding photons within a
given percentile of the point spread function of the instrument from a
potential source.  Furthermore, we will assume a simple expression for
the likelihood that assumes that the photons from within a particular
region of the sky come from the combination of a point source at the
centre of the region and a flat background.
We do not assume a particular spectral model but only that the
measured momenta of photons from a point source follow the expected
distribution from a point source as determined by measurements and
simulations \citep{2013ApJ...765...54A}.

\section{The Photon Database}
\label{sec:photon-database}

The key to the speed of this algorithm is the database that contains
the positions of the observed photons on the sky.  Each photon is
stored in a four-dimensional $k-d$~tree
\citep{Bentley:1975:MBS:361002.361007}.  A $k-d$~tree is a binary tree
in which each successive branching splits the space in two regions
along a hyperplane perpendicular to the last splitting.  For example,
to construct an efficient two-dimensional $k-d$~tree, we would first
find the median $x$-coordinate of our points and assign it to be the
root of the tree.  The left branch would contain all the points with
$x$ values less than the median, and the right branch would contain
all the points with $x$ values greater than the median.  Now we repeat
the process with only the points on the left branch and use the
$y$-coordinate instead of the $x$-coordinate, so we now have
left-upper branch with $x$-values less than the median $x$-value and
$y$-values greater than the median $y$-value for the points on the
left branch.  Similarly, we have a left-lower branch, and we repeat
the process on the right branch, generating a right-upper and
right-lower branches.  The next step in the two-dimensional $k-d$~tree
is to repeat on each of the branches with the $x-$coordinate again.
The branching points do not have to be the precise medians of the
data; one can construct well-balanced and efficient trees by sampling
a fraction of the points to estimate the medians.  The process
continues until all the branches end with single points.  Our photon
data is four-dimensional.  We use $x$, $y$ and $z$ to represent a
photon's position on the celestial sphere and $w$ to encode the
95\%-energy-containment radius. Therefore, in our case, we construct a
four-dimensional $k-d$~tree, so we split the data along a plane with
constant $x-$coordinate, constant $y$, constant $z$, constant $w$ and
then repeat with the $x-$coordinate.  Once a $k-d$~tree is constructed
for the data, one can search for points in our case photons within a
given range of a particular position on the sky efficiently.  Because
the points are arranged in a binary tree, finding the photons that
arrived within a given angle of a particular direction takes only on
order of $\log_2 N$ operations where $N$ is the total number of
photons in the database.

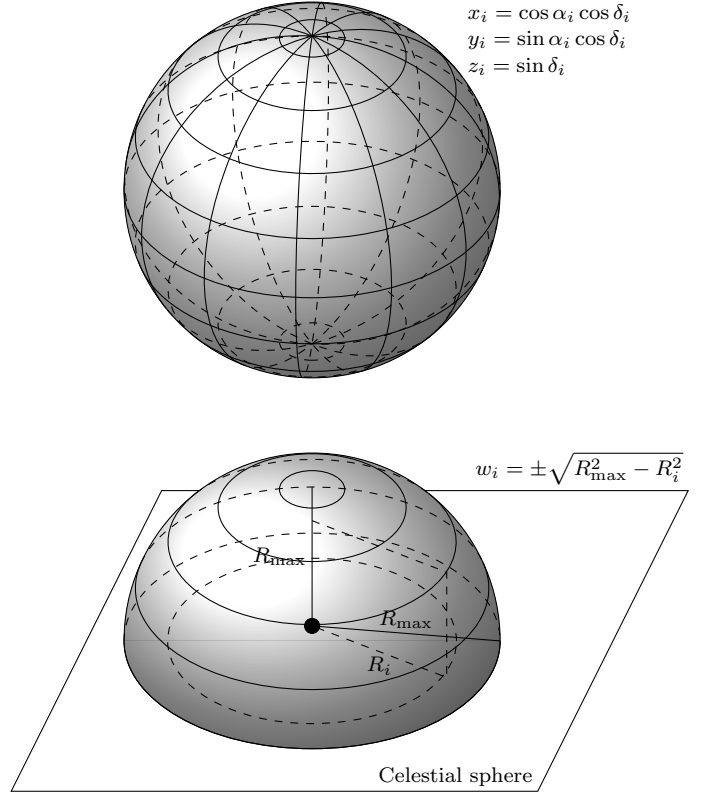
\begin{figure}
\begin{tikzpicture} 

\def\R{2.5} 
\def\angEl{35} 
\filldraw[ball color=white] (0,0) circle (\R);
\foreach \t in {-80,-60,...,80} { \DrawLatitudeCircle[\R]{\t} }
\foreach \t in {-5,-35,...,-175} { \DrawLongitudeCircle[\R]{\t} }
\draw (2,2.35) node [right] {$x_i=\cos\alpha_i\cos\delta_i$} 
      (2,2.0) node [right] {$y_i=\sin\alpha_i\cos\delta_i$}
      (2,1.65) node [right] {$z_i=\sin\delta_i$};
\begin{scope}[shift={(0,-6)}]
\draw (-4,-2) -- (-2,2) -- (5,2) -- (3,-2) -- cycle;
\draw (3,-2) node [above left] {Celestial sphere};
\filldraw[ball color=white] (0:\R) arc (0:180:\R)
(0,0) (0:\R) arc (0:-180:2.5 and 1.433904196);
\foreach \t in {0,20,...,80} { \DrawLatitudeCircle[\R]{\t} }
\draw (2.1,2.3) node [right] {$w_i=\pm\sqrt{R^2_\mathrm{max}-R^2_i}$};
\draw (0,0.2) -- node [above] {$R_\mathrm{max}$} (\R,0)
      (0,0.2) -- node [left] {$R_\mathrm{max}$} (0,0.81916*\R);
\draw [dashed] (0,1.6) -- ++(1.79,-0.6855) -- ++(0,-1.4) 
      -- node [below] {$R_i$} (0,0.2)
      (0,0.0) circle (1.92 and 1.10);
\filldraw[black] (0,0.2) circle (0.1);
\end{scope}
\end{tikzpicture}
\caption{The location of a given photon event on the celestial sphere
  and in the additional dimension. $R_i$ is the ninety-fifth
  percentile radius for the photon in question and $R_\mathrm{max}$ is
  the largest ninety-fifth percentile radius for the photons in the
  sample.}
\label{fig:sphere}
\end{figure}

We use the particularly memory efficient implementation of
\citet{LangPhD} (used in astrometry.net).  The coordinates are
actually stored as shorts instead of floats to save additional memory.
The typical coordinates range from $-1$ to $+1$, so using shorts
yields an angular precision of about six arcseconds much finer than
that of the Fermi point-spread function (PSF). This memory efficient
implementation allows us to store all the photons detected by Fermi
above 100~MeV and within a zenith angle of 100 degrees in memory
simultaneously.  The first three dimensions contain the position of
the photon on the celestial sphere as shown in the upper portion of
Fig.~\ref{fig:sphere}.  Storing the direction of the photon momentum
in this manner removes the coordinate singularity of the spherical
coordinates.
The fourth
coordinate that we denote by $w$ depends on the point-spread function
for the photon in question.  In particular $w=\pm
\sqrt{R^2_\mathrm{max}-R_i^2}$ where $R_i$ is the radius of the
ninety-fifth percentile at the energy, entrance angle and front or
back conversion for the photon.  The choice of the ninety-fifth
percentile is arbitrary.  Choosing a smaller cut-off radius would
decrease the number of photons from the candidate point source and also
from the background.  For convenience, we use positive values of $w$
for front-converted photons and negative values of $w$ for
back-converted photons.  Furthermore, $R_\mathrm{max}$ is the largest
value of the ninety-fifth percentile of all the photons in the sample,
{\em i.e.} the ninety-fifth percentile for the photon with the poorest
angular resolution.  This is depicted in the lower panel of
Fig.~\ref{fig:sphere}.

\begin{figure}
\includegraphics[width=\columnwidth]{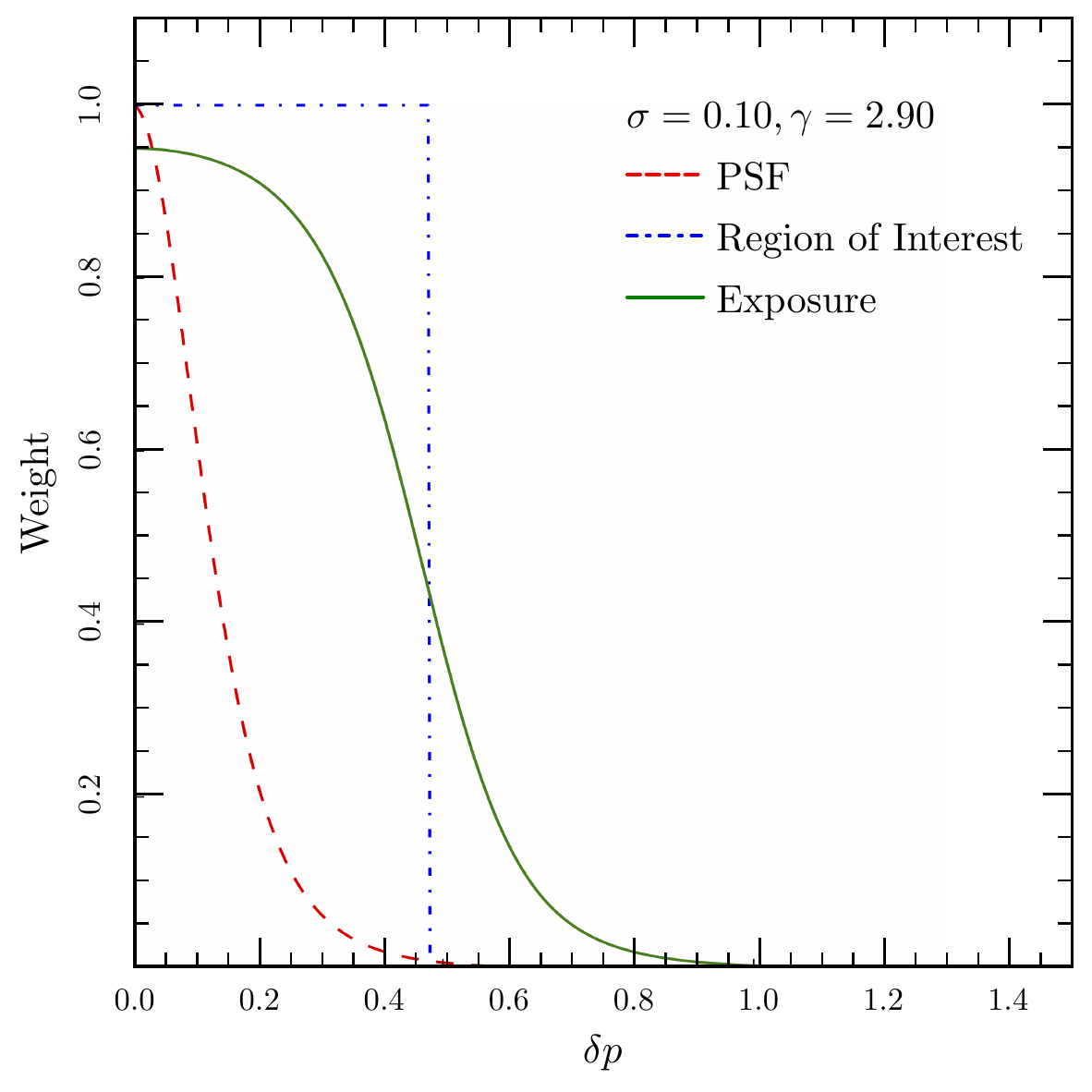}
\caption{The Exposure Map, Region of Interest (ROI)
  and Point-Spread Function (PSF) with $\sigma$ and $\gamma$ characterise
  the width of the King function and its asymptotic power-law dependence respectively
  according to Eq.~\ref{eq:1}.
}
\label{fig:expmap}
\end{figure}

Once the $k-d$~tree is created, it is efficient to find all the
entries within the database within a given Cartesian distance of a
particular point.  In our case we query the database for all the
photons within a distance $R_\mathrm{max}$ of a particular point on
the celestial sphere and use $w=0$ for the fourth coordinate.  The
particular choice of $w$ for the observed photons means that the query
will yield all the photons that are within the ninety-fifth
percentile of the PSF.  In other words, if there is a point source
located at that particular point the query will return on average 95\%
of the photons from that source.  Of course, it will also return
photons from the background and other nearby sources.  Additionally,
the region of interest for a particular prospective source is energy
dependent.  The form of the exposure map also is energy dependent as
shown in Fig.~\ref{fig:expmap}.  It peaks at 0.95 times the exposure
time in the direction of the potential source and slowly drops to the
ninety-fifth percentile of the PSF in radius and then drops according
to the power-law of the tail component of the linear combinations of
King functions,
\begin{equation}
  K(\delta p; \sigma, \gamma) =\frac{1}{2\pi \sigma^2} \left ( 1- \frac{1}{\gamma} \right ) \left [ 1 + \frac{1}{2\gamma} \left (\frac{\delta p}{\sigma} \right )^2 \right ]^{-\gamma}
\label{eq:1}
\end{equation}
that are used to characterise the Fermi PSF \citep{2013ApJ...765...54A}.  

Although the construction of the tree is not done in parallel, the
queries are performed in parallel using the tree held in shared
memory.  For example if one uses all the photons above 100~MeV from
weeks 9 through 216 with a standard zenith angle cut of less than 100
degrees (89,684,009 photons) requires one gigabyte to store the tree
and about ten minutes to construct. We perform 200,000 location
queries corresponding to a HEALPix \citep{2005ApJ...622..759G} grid
with $\mathtt{NSIDE}=128$ (see \S~\ref{sec:source-likelihood} for
further details).  The 200,000 location queries and likelihood
calculations require 140,000 seconds (700~ms each), so the speed-up
through parallisation can be dramatic.  On the other hand, if one
restricts to photons above 1~GeV (13,193,171), it only takes 90
seconds to construct the tree.  At higher energies, it makes sense to
make more location queries because the PSF is smaller.  In this
example 786,426 queries require 5,600 seconds of CPU time (7~ms each)
or only six minutes on sixteen cores.

\section{Source Likelihood}
\label{sec:source-likelihood}

Using the $k-d$~tree to determine the photons that lie within the 95\%
energy enclosure region of the point-spread function, we calculate
several statistics of the observed photons to assess the likelihood of
a source being at a particular position on the sky.  We first
determine two statistics whose distributions are potentially
known. First, if all the photons within the region of interest (all
photons that lie within the 95th-percentile cone of the potential
source) indeed come from a uniform background, we would expect that
the photons would be uniformly distributed in the region of interest.
Therefore, in this case the ratio of the solid angle enclosed in a
circle centred on the potential source and running through the
observed photon to the total solid angle within the region of interest
for that particular photon should be uniformly distributed between
zero and one. This is just a $V/V_\mathrm{max}$-test
\citep{1968ApJ...151..393S} adapted to the distribution of the photons
on the sky.

We denote the mean of this ratio over the
observed photons $\bar r^2$, that is
\begin{equation}
  \bar r^2 = \sum_i \left ( \frac{r_i}{r_{\mathrm{max},i}} \right )^2 
\end{equation}
where $r_i$ is the angular distance between the photon direction and
the direction of the potential source and $r_{\mathrm{max},i}$ is the
95\% energy containment radius for the photon.  This statistic is
distributed as a Bates distribution with mean of $1/2$ and variance of
$1/(12 N_\gamma)$.  

Second, if all the photons within the
region of interest indeed come from a point source at the centre of
the region of interest that extends to the ninety-five percentile, the
ratio of the percentile of a given photon within the cumulative PSF
distribution for that particular photon to 0.95 should be uniformly
distributed between zero and one.  For convenience because we only
probe the photons out to the 95\% energy containment radius, we define
the cumulative fraction of energy out to the ninety-fifth percentile
\begin{equation}
  C_\mathrm{PSF} (r) = \frac{1}{0.95} \int_0^r \mathrm{PSF}(r^\prime) 2 \pi r^\prime d r
  \label{eq:2}
\end{equation}
and 
\begin{equation}
  \bar f = \sum_i C_\mathrm{PSF} (r_i).
  \label{eq:3}
\end{equation}
This will be distributed as a Bates distribution if all of the photons
come from a point source.

If we assume that the observed photons originate from a linear
combination of these two possiblities, we can determine the ratio of
the two contributions from these statistics.  For $\bar r^2$ we have
\begin{eqnarray}
  \bar r^2 &=& A_f \int_0^1 \left (\frac{r}{r_\mathrm{max}} \right )^2 d \left (C_\mathrm{PSF} \right ) + \label{eq:4} \\
  & & ~~~ 
   (1 - A_f) \int_0^1 \left (\frac{r}{r_\mathrm{max}} \right )^2 d \left (\frac{r}{r_\mathrm{max}} \right )^2 \nonumber
\end{eqnarray}
and for $\bar f$ we have
\begin{eqnarray}
  \bar f &=& A_f \int_0^1 C_\mathrm{PSF}  d \left (C_\mathrm{PSF} \right ) + \label{eq:5} \\
  & & ~~~  (1 - A_f) \int_0^1 C_\mathrm{PSF} d \left (\frac{r}{r_\mathrm{max}} \right )^2.
    \nonumber
\end{eqnarray}
To move forward we examine the first term in $\bar r^2$
\begin{eqnarray}
  B &=& \int_0^1 \left (\frac{r}{r_\mathrm{max}} \right )^2 d \left (C_\mathrm{PSF} \right )
  = \left . \left (\frac{r}{r_\mathrm{max}} \right )^2 C_\mathrm{PSF} \right |_0^1 - \\
  \label{eq:6}
  & & ~~~  \int_0^1 C_\mathrm{PSF} d \left (\frac{r}{r_\mathrm{max}} \right )^2. \nonumber \\
  &=& 1 - \int_0^1 C_\mathrm{PSF} d \left (\frac{r}{r_\mathrm{max}} \right )^2
  \label{eq:7}
\end{eqnarray}
where we have integrated the first term by parts to get one minus the
second term in $\bar f$.  We can substitute these results into
Eq.~\ref{eq:4} and~\ref{eq:5} to yield
\begin{eqnarray}
  \bar r^2 &=& A_f B + \frac{1-A_f}{2} \label{eq:8}
  \\
  \bar f^2 &=& \frac{A_f}{2} + (1 - A_f) (1 - B) \label{eq:9}
\end{eqnarray}
Using this result, we can solve for the fraction of photons that come
from the point source to yield
\begin{equation}
  A_f=\frac{\frac{1}{2}-\bar r^2}{\bar f-\bar r^2}.
  \label{eq:10}
\end{equation} 
We can also find the value of $B$ to be
\begin{equation}
  B = \bar r^2 - \bar f + \frac{1}{2}.
  \label{eq:11}
\end{equation}
The estimator $A_f$ is calculated from various statistics of the
sample and assumes that the photons come from a single point source
and a flat background, so it can have sampling errors that cause it to
be out of the range of 0 to 1, and systematics errors as well
({\em e.g.} the photons come from multiple sources) that can also put it out
of this range. We can estimate the significance of the value of $A_f$ by
\begin{equation}
  S(r^2) = \left ( \bar r^2-\frac{1}{2} \right ) \sqrt{12 N_\gamma}.
\label{eq:12}
\end{equation}
We have retained the sign of the deviation from the source-free value,
so that a negative value of $S(r^2)$ indicates that the photons within
the region of interest are centrally concentrated.  The probabilty of
getting a value of $|S(r^2)|$ larger than $x$ by chance is
\begin{equation}
  P\left [ |S(r^2)| > x \right ] = \frac{1}{2} \mathrm{erfc} \left ( \frac{x}{\sqrt{2}} \right
  ) \approx \exp \left (-\frac{x^2}{2} \right )
  \label{eq:13}
\end{equation}
if we take the limit of many photons in the region of interest where
the Bates distribution tends to the normal distribution.  

\begin{table*}
  \caption{Basic statistics calculated for the photon distribution around a potential source}
  \label{tab:stats}
  \begin{tabular}{l|cll}
    \hline
    Statistic & Symbol & Abbreviation & Definition \\
    \hline
    Number of photons                     &  $N_\gamma$ & \texttt{N}        & Number of photons that lie within the 95\% percentile \\
    Mean Solid Angle Ratio                &  $\bar r^2$          & \texttt{MEANR2}   & The mean of the ratio of solid angle enclosed between observed \\
                                          &                      &                   & photon position and source location and the solid angle \\
                                          &                      &                   & enclosed with the 95\% percentile of the PSF \\
    Mean Percentile Ratio                 &  $\bar f$            & \texttt{MEANFRAC} & The mean of the ratio of PSF percentile to 95\% \\
    Significance of \texttt{MEANR2}       & $S(r^2)$             & \texttt{SIGR2}    & How many standard deviations is \texttt{MEANR2} away from 0.5 \\
    Significance of \texttt{MEANFRAC}     & $S(f)$               & \texttt{SIGFRAC}  & How many standard deviations is \texttt{MEANFRAC} away from 0.5 \\
    Fraction of photons from point source & $A_f$                & \texttt{AFRAC}    & If one assumes that the photons come from the sum of a  \\
                                          &                      &                   & uniform background and a point source, what fraction \\
                                          &                      &                   & come from the point source? $A_f=(0.5-\bar r^2)/(\bar f-\bar r^2)$ \\
    Amplitude of PSF in Likelihood Fit      & $A_\mathrm{PSF}$      & \texttt{APSF}     & Fraction of photons from the region of interest that come \\
                                          &                      &                   & from the source according to the likelihood fit \\
  \end{tabular}
\end{table*}

These basic statistics are summarized in Tab.~\ref{tab:stats}, and
Tab.~\ref{tab:topten} lists these statistics for the fifteen most
significant sources detected.  These basic statistics really just
compare two numbers about the distribution of the photons within the
region of interest.  We can use the detailed knowledge of the point
spread function to develop a more comprehensive test of the
distribution of photons.  In particular, we define the unbinned
likelihood \citep[this is very similar to the expression used by][for neutrino telescopes]{2008APh....29..299B}
\begin{equation}
  \log L = \sum_\gamma \log \left [ A_\mathrm{PSF} \frac{\mathrm{PSF}_i \Omega_{\mathrm{max},i} }{0.95} + (1
    - A_\mathrm{PSF}) \right ]
  \label{eq:14}
\end{equation}
where we have dropped $N_\mathrm{pred}$ from the usual definition
because we have defined the model in such a way that
$N_\mathrm{pred}=N_\gamma$ automatically and $d
N_\mathrm{pred}/dA_\mathrm{PSF}=0$.  The variable
$\Omega_\mathrm{max,i}$ is the solid angle that encloses 95\% of
energy for the PSF of that particular photon, and $\mathrm{PSF}_i$
is the value of the PSF for the photon if it indeed comes from the
position of the candidate source.  The maximum of $\log L$ is found
by numerically varying the value $A_\mathrm{PSF}$.  For
$A_\mathrm{PSF}=0$, $\log L=0$ and because we are fitting a single
variable, $\log L$ is distributed as a chi-squared distribution with
a single degree of freedom and the probability of getting a value of
$\log L$ larger than $x$ by chance is
\begin{equation}
  P(\log L > x) = \sqrt{\pi} \mathrm{erfc} \left ( \sqrt{\frac{x}{2}}
  \right ) \approx \exp \left (-\frac{x}{2} \right ).
  \label{eq:15}
\end{equation}
From Tab.~\ref{tab:topten} we can see that the values of $A_f$ are
similar to those of $A_\mathrm{PSF}$ at least for highly significant
sources.

The first pass is to determine the value of $\log L$ on a HEALPix grid
\citep{2005ApJ...622..759G} of potential sources.  The HEALPix grid is
an array of equal area regions on a sphere, in this case the sky.  The
sphere is first projected onto a rhombic dodecahedron.  Each of the
faces is a rhombus congruent to the others, and a pair of opposite
vertices is aligned with the poles.  Each of the rhombuses is
subdivided into equal solid angle regions by dividing in each
dimensional in two.  The parameter $\mathtt{NSIDE}$ is the number of
subdivisions and must be a power of two, so the total number of
regions is $12 \times \mathtt{NSIDE}^2$.

In the FermiFAST algorithm we use the HEALPix grid for the initial
guesses for source positions and to find local maxima on the grid.
Once the approximate local maxima are found on the grid, we refine the
position of the detection to increase the likelihood further.
Consequently, the initial spacing of the grid should be smaller than
the angular resolution of the LAT at the energy range of interest, so
that separate sources result in separate peaks in the initial
likelihood map.  Otherwise, the value of $\mathrm{NSIDE}$ is
arbitrary.  For example for the photons above 1~GeV we used
$\mathtt{NSIDE}=256$ or 786,432 grid points, so the grid points lie
approximatly 13 arcminutes apart or about twenty points per square
degree and 60 measurements within the typical 68\% energy containment
radius at 1~GeV \citep{2012ApJS..203....4A}. Of these 786,432 points,
18,425 have $P(\log L >x) < e^{-12.5} \approx 4 \times 10^{-6}$.  This
is a five-sigma threshold. Next we take this list of potentially
significant sources and find the local maxima of $\log L$; this
reduces the number to 1,226 unique potential sources.  However, 426
have $A_\mathrm{PSF}<0$, indicating not a source but a point-like
deficit, leaving 800 sources.  Each of these potential source
positions is used as an initial guess to optimize the local maxima of
$\log L$ and get a more precise position for each source.  For the
lower-energy cutoff, we start with $\mathtt{NSIDE}=128$ or 196,608
grid points or a separation of 26 arcminutes.  Of these 196,608
points, 96,066 have $P(\log L >x) < e^{-12.5} \approx 4 \times
10^{-6}$ and 967 unique peaks of which 310 have $A_\mathrm{PSF}>0$.

\begin{figure*}
  \includegraphics[width=\columnwidth]{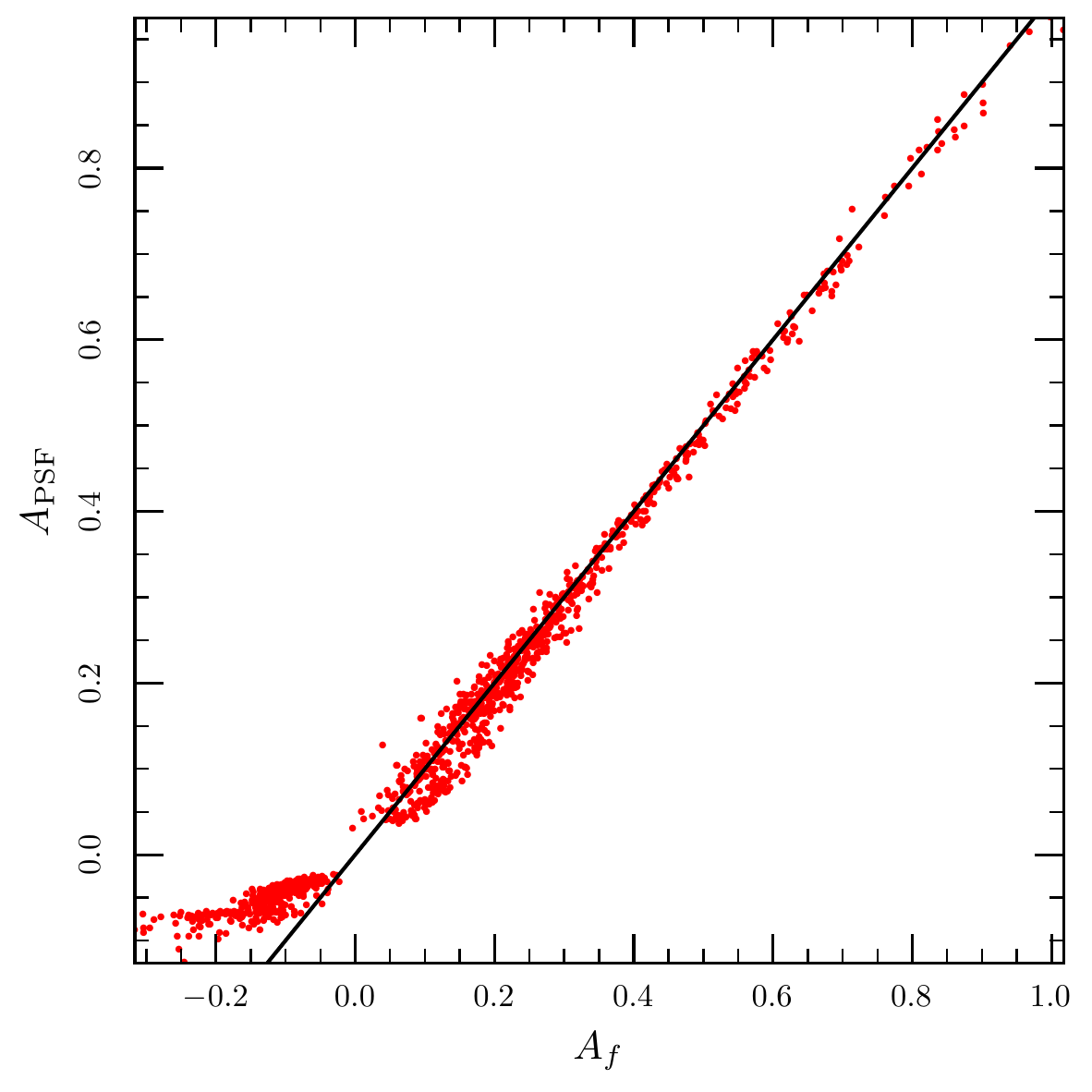}
  \includegraphics[width=\columnwidth]{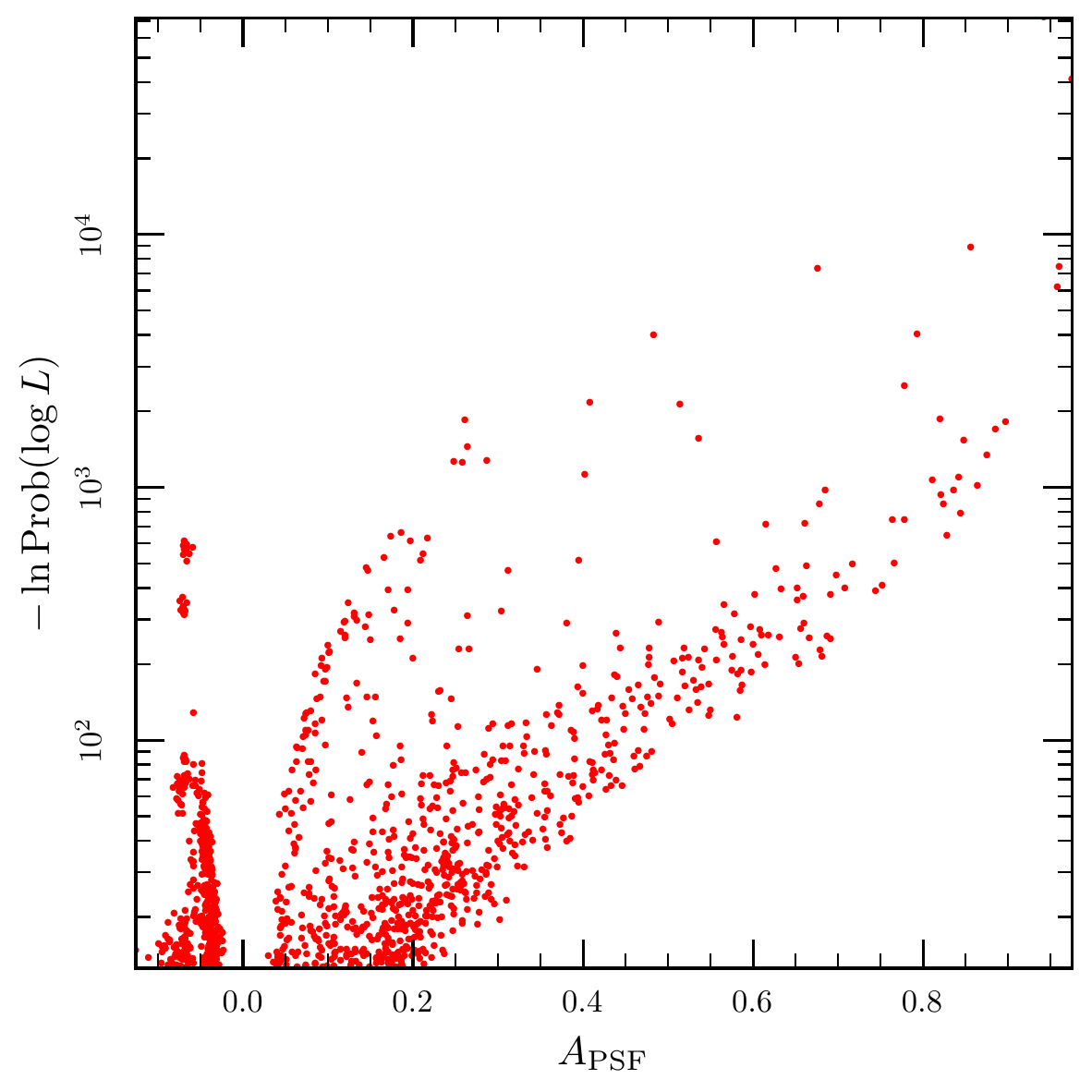}
  \caption{Left: the correlation of the value $A_f$, determined from
    the means of the photon distributions, to the value of
    $A_\mathrm{PSF}$, determined from the maximum likelihood
    techinque.  The line traces $A_\mathrm{PSF}=A_f$. Right: The
    values of $A_\mathrm{PSF}$ against the significance.}
  \label{fig:acorr}
\end{figure*}

Before discussing the results ({\em i.e.} the catalogue of sources) and
how they compare with the third Fermi catalogue, we will examine the
performance of the technique and focus on the sources detected from
photons with energies exceeding 1~GeV.  The left-panel of
Fig.~\ref{fig:acorr} depicts the results of the initial set of peaks
with $P(\log L)<e^{-12.5}$ before position refinement.  We see that
for $A_f>0$ there is a strong linear correlation between the value of
$A_f$ and $A_\mathrm{PSF}$.  Whereas for negative values of $A_f$
where there is a hole in the photon distribution, the value of
$A_\mathrm{PSF}$ saturates at $-0.1$. The right panel shows the values
of $A_\mathrm{PSF}$ against the significance.  There are both
significant sources and holes in the photon distribution.  The
fractional deficit in the ``hole'' regions is limited to one tenth.
Because of the focus of this paper is to look for gamma-ray sources,
we will not discuss these ``holes'' further, but they would be an
interesting focus of further investigation. 

\begin{figure*}
  \includegraphics[width=\columnwidth]{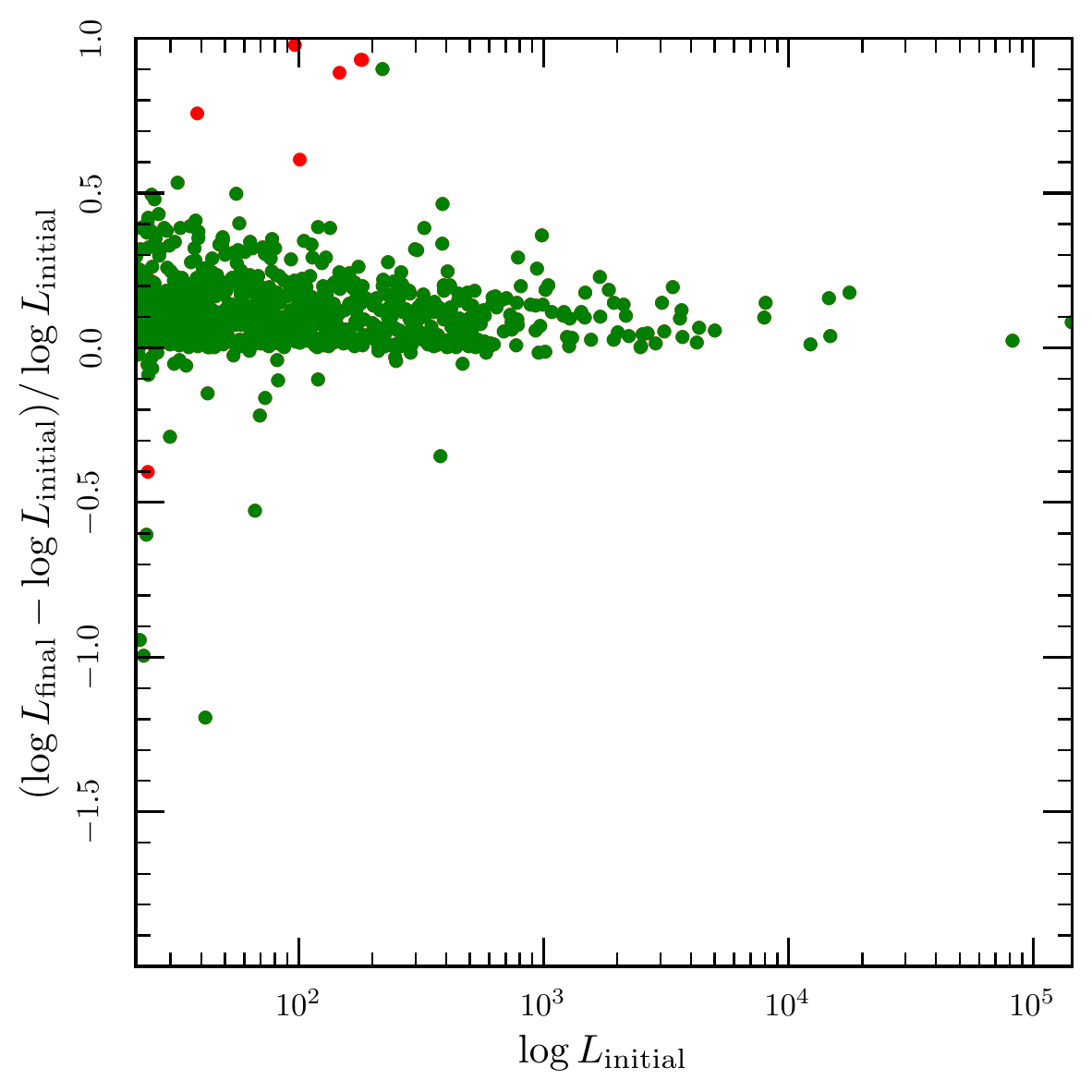}
  \includegraphics[width=\columnwidth]{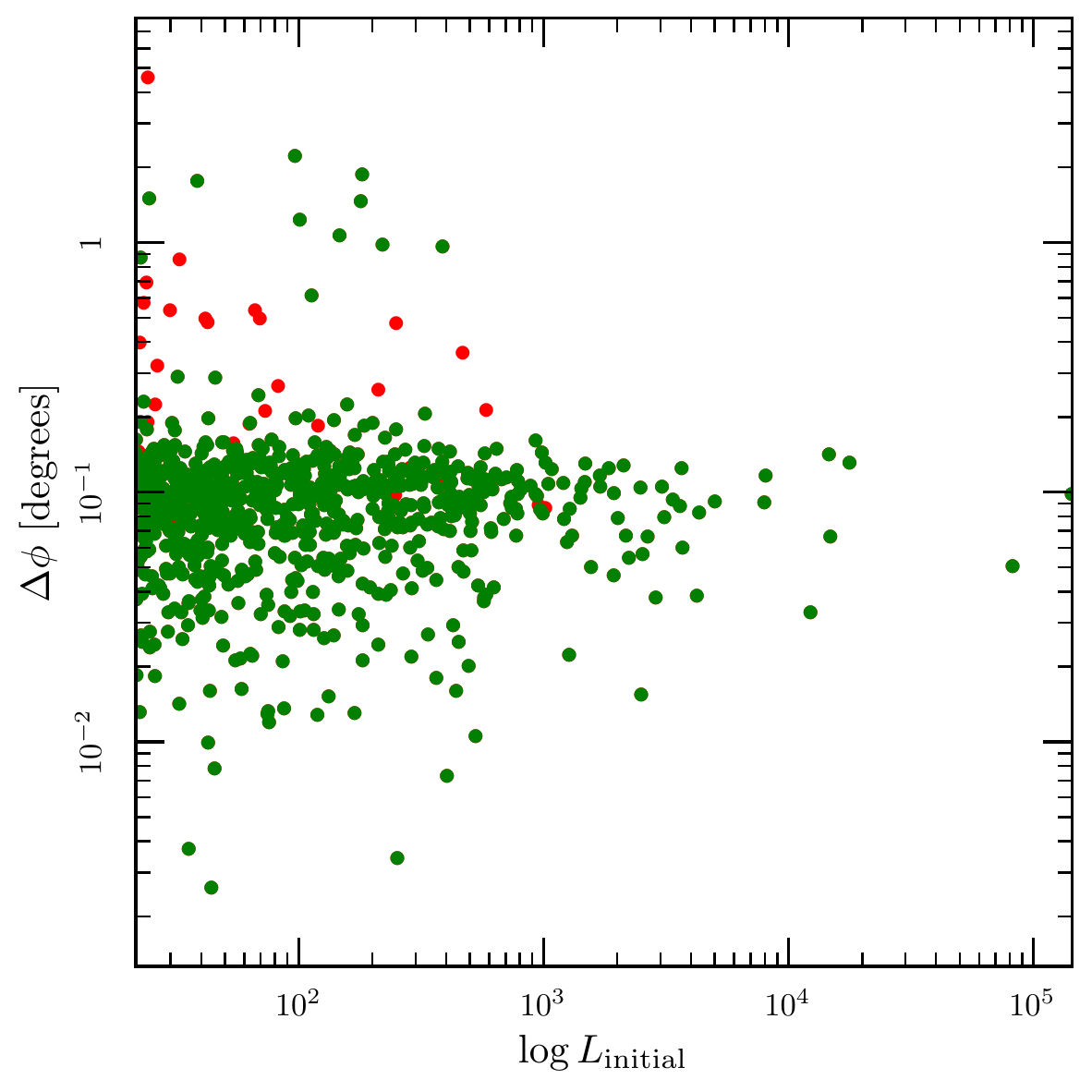}
  \caption{Left: the relative change in $\log L$ during the position
    refinement.  Green is for sources whose positions moved less than
    one degree.  Red is for greater movement.  Right: the relative
    change in position during the position refinement.  Green is for
    sources whose positions likelihood increased.  Red is for those
    whose likelihood decreased.}
  \label{fig:position}
\end{figure*}

For the sources in the preliminary catalogue ({\em i.e.} those with
$A_\mathrm{PSF}>0$) we further refine the position estimate of the
source.  The left panel of Fig.~\ref{fig:position} depicts the change
in $\log L$ during the optimization.  Sometimes the value of $\log L$
actually decreases, but usually it increases modestly by say ten
percent.  The right panel shows the change in the position of the
source.  The size of each HEALPix region is about 0.2 degrees on a
side, so an optimization within a given HEALPix cell would result in a
typical movement of about one tenth of a degree, and the results bear
this out.  Occasionally the estmated source position moves by a
large distance indicating that the optimization has found another
nearby peak.  We use the optimized position only if the value of $\log
L$ has actually increases, and the estimated position of the source
has shifted by less than 0.5 degrees.

\section{Results}
\label{sec:results}

To test the algorithm we used nearly the same data set as used to
construct the Fermi Large Area Telescope Third Source Catalog
\cite[3FGL][]{2015ApJS..218...23A}.  We used weeks 9 through 216. That
is, photons detected between 2008 August 4 (15:45:36 UTC) and 2012
July 26 (01:07:25 UTC), a span of nearly four years.  This is about
five days shorter than the span of the 3FGL observations because we
used the weekly files.  We used the good-time intervals (GTI) as
defined by the Fermi team, the Pass 7 response function
(\texttt{P7REP\_SOURCE\_V15}) and the Pass 7 reprocessed data.  We use
the Pass 7 data for comparison with the analysis of the 3FGL which
also used Pass 7 data. From the form of the likelihood function,
Eq.~\ref{eq:14}, it is apparent that we do not use a model for the
background, and the key ingredient of the instrument response is the
estimate of the point-spread function \citep{2012ApJS..203....4A}.

We create two catalogues: one using photons above 1~GeV and the second
with all photons above 100~MeV.  The fifteen most significant sources in
the 1~GeV map are given in Tab.~\ref{tab:topten} with the refined
positions and the original detection significances.  We see a general
trend that the significances in sigma-units of the 3FGL are about
three times that of the FermiFAST technique.  The two exceptions in
the table are the Crab pulsar and 3FGL J1745.6-2859c/3FGL
J1745.3-2903c.  The Fermi analysis splits the Crab source among three
sources: the pulsar itself, the synchrotron and inverse Compton
components of the pulsar wind nebulae.  The FermiFAST source most
closely coincides with the pulsar, and the 3FGL significance for this
component alone is quoted in the table.  The thirteenth FermiFAST
source lies close to both 3FGL J1745.6-2859c (1.8 arcminutes) and 3FGL
J1745.3-2903c (7.1 arcminutes).  It is closer to 3FGL
J1745.6-2859c which has a lower 3FGL significance than the other
potential counterpart.
%
%

\begin{table*}
  \caption{The results for the fifteen most significant peaks in the
    1~GeV map.  We list two entries for the thirteenth source because
    there are two 3FGL sources within 6 arcminutes of each and the
    FermiFAST source.  The closer source to the FermiFAST detection is
    listed first.}
  \label{tab:topten}
  \begin{tabular}{lrrrrrrrrrrr}
    \hline
     Source & \multicolumn{1}{c}{RA} & \multicolumn{1}{c}{Dec}  & \multicolumn{1}{c}{$N_\gamma$}  &\multicolumn{1}{c}{$\bar r^2$} & \multicolumn{1}{c}{$\bar f$} & \multicolumn{1}{c}{$S(r^2)$} & \multicolumn{1}{c}{$S(f)$} & \multicolumn{1}{c}{$A_f$}  & \multicolumn{1}{c}{$A_\mathrm{PSF}$} & \multicolumn{1}{c}{$S(\mathrm{FF})$} & \multicolumn{1}{c}{$S(\mathrm{3FGL})$} \\
    \hline

PSR J0835-4510 (Vela)     & $128.84$ & $-45.18$ & 171821 & $0.174$ & $0.521$ & $-467.5$ & $  29.5$ & $0.94$  & $0.94$ & $   378.92$ & $  1048.96$  \\ 
PSR J0633+1746 (Geminga)  & $ 98.48$ & $ 17.77$ &  90467 & $0.164$ & $0.501$ & $-349.9$ & $   0.6$ & $1.00$  & $0.97$ & $   286.73$ & $  1012.14$  \\ 
PSR J0534+2200 (Crab)     & $ 83.64$ & $ 22.02$ &  26443 & $0.204$ & $0.558$ & $-166.7$ & $  32.6$ & $0.84$  & $0.86$ & $   133.34$ & $    30.67$  \\ 
  LAT PSR J1836+5925      & $279.06$ & $ 59.43$ &  16595 & $0.167$ & $0.494$ & $-148.8$ & $  -2.5$ & $1.02$  & $0.96$ & $   121.85$ & $   438.12$  \\ 
      PSR J1709-4429      & $257.42$ & $-44.48$ &  32526 & $0.265$ & $0.614$ & $-146.9$ & $  71.3$ & $0.67$  & $0.68$ & $   121.02$ & $   360.82$  \\ 
            3C 454.3      & $343.50$ & $ 16.15$ &  14021 & $0.170$ & $0.511$ & $-135.4$ & $   4.4$ & $0.97$  & $0.96$ & $   110.98$ & $   480.74$  \\ 
  LAT PSR J0007+7303      & $  1.76$ & $ 73.05$ &  13558 & $0.224$ & $0.564$ & $-111.5$ & $  25.6$ & $0.81$  & $0.79$ & $    89.89$ & $   288.75$  \\ 
  LAT PSR J2021+4026      & $305.39$ & $ 40.45$ &  31304 & $0.331$ & $0.669$ & $-103.8$ & $ 103.9$ & $0.50$  & $0.48$ & $    89.31$ & $   237.05$  \\ 
      PSR J1057-5226      & $164.49$ & $-52.46$ &   8194 & $0.230$ & $0.570$ & $ -84.6$ & $  21.8$ & $0.80$  & $0.78$ & $    70.87$ & $   200.06$  \\ 
      PSR J2021+3651      & $305.26$ & $ 36.85$ &  22683 & $0.355$ & $0.693$ & $ -75.5$ & $ 100.5$ & $0.43$  & $0.41$ & $    65.78$ & $   145.14$  \\ 
   LS I+61 303            & $ 40.14$ & $ 61.22$ &  14060 & $0.323$ & $0.667$ & $ -72.6$ & $  68.5$ & $0.51$  & $0.51$ & $    65.12$ & $   196.61$  \\ 
   PKS 1510-08            & $228.21$ & $ -9.10$ &   5718 & $0.217$ & $0.555$ & $ -74.0$ & $  14.5$ & $0.84$  & $0.82$ & $    60.86$ & $   207.27$  \\ 
   3FGL J1745.6-2859c     & $266.45$ & $-28.99$ &  48087 & $0.388$ & $0.750$ & $ -85.3$ & $ 189.8$ & $0.31$  & $0.26$ & $    60.60$ & $    11.97$  \\
   3FGL J1745.3-2903c     &          &          &        &         &         &          &          &         &        &             & $    20.66$  \\
   PKS 0537-441           & $ 84.71$ & $-44.09$ &   4692 & $0.188$ & $0.534$ & $ -74.1$ & $   8.1$ & $0.90$  & $0.90$ & $    60.15$ & $   219.16$  \\ 
   Mkn 421                & $166.12$ & $ 38.21$ &   4527 & $0.187$ & $0.545$ & $ -73.0$ & $  10.5$ & $0.87$  & $0.89$ & $    58.16$ & $   190.35$  \\ 

\end{tabular}
\end{table*}

\begin{figure*}
  \includegraphics[width=\textwidth,clip,trim=0.6in 1in 0.8in
    1in]{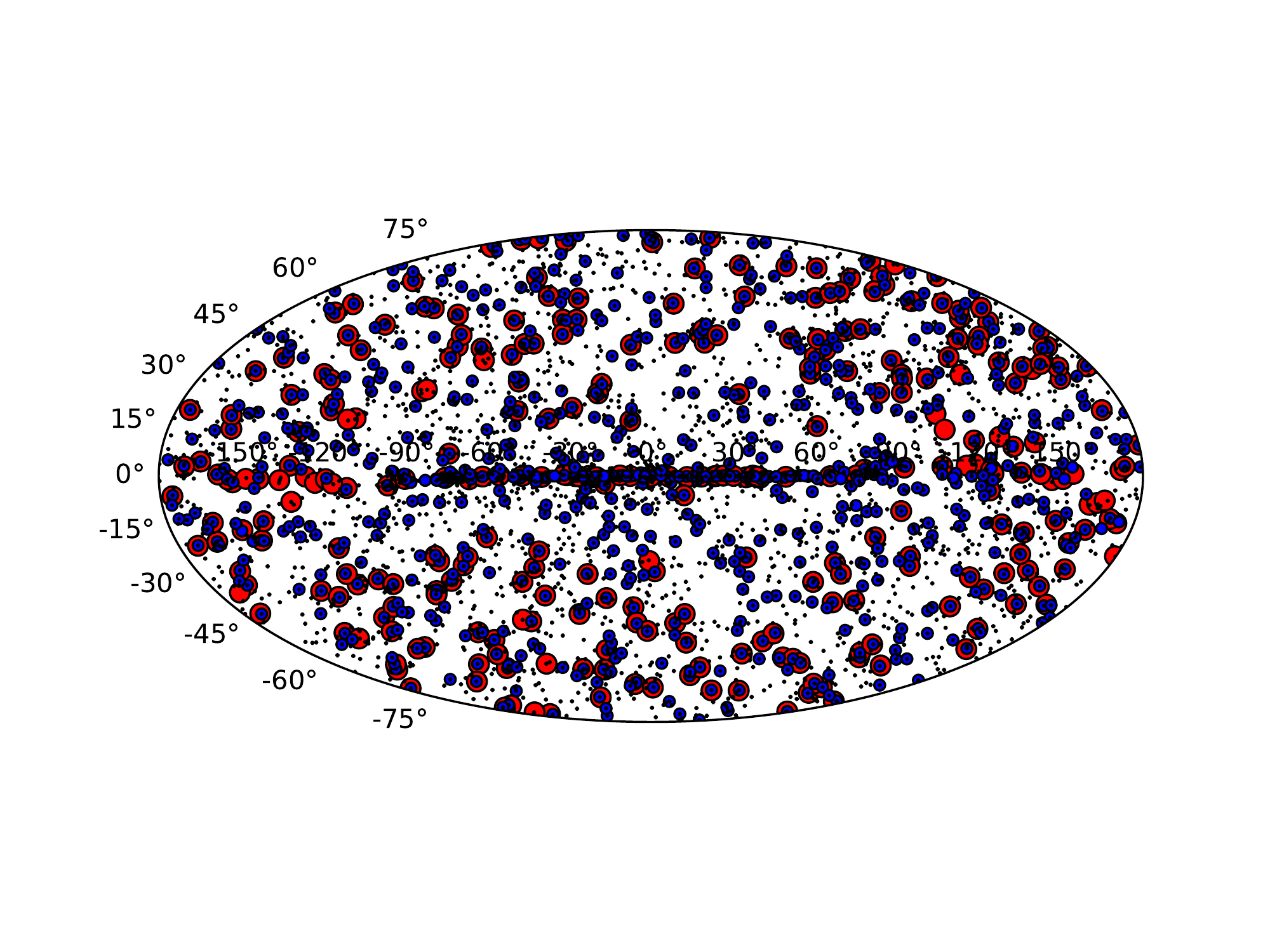}
  \caption{Sources with $\ln P(\log L)<-12.5$ (i.e. five sigma) in the
    1~GeV catalogue (small blue circles), in the 100~MeV catalogue
    (big red circles) and the 3FGL (black dots).}
  \label{fig:sources}
\end{figure*}

\begin{figure*}
\includegraphics[width=\columnwidth]{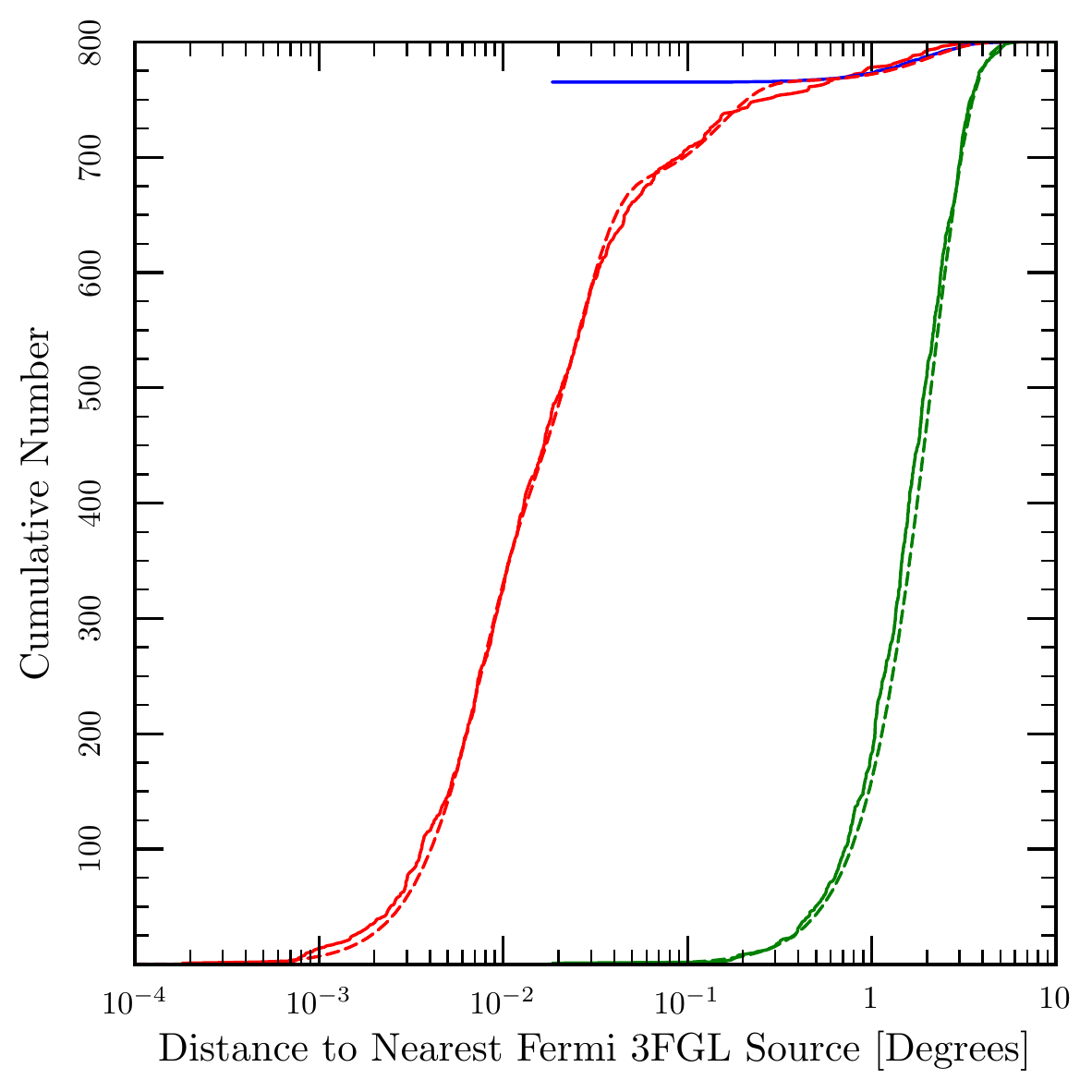}
\includegraphics[width=\columnwidth]{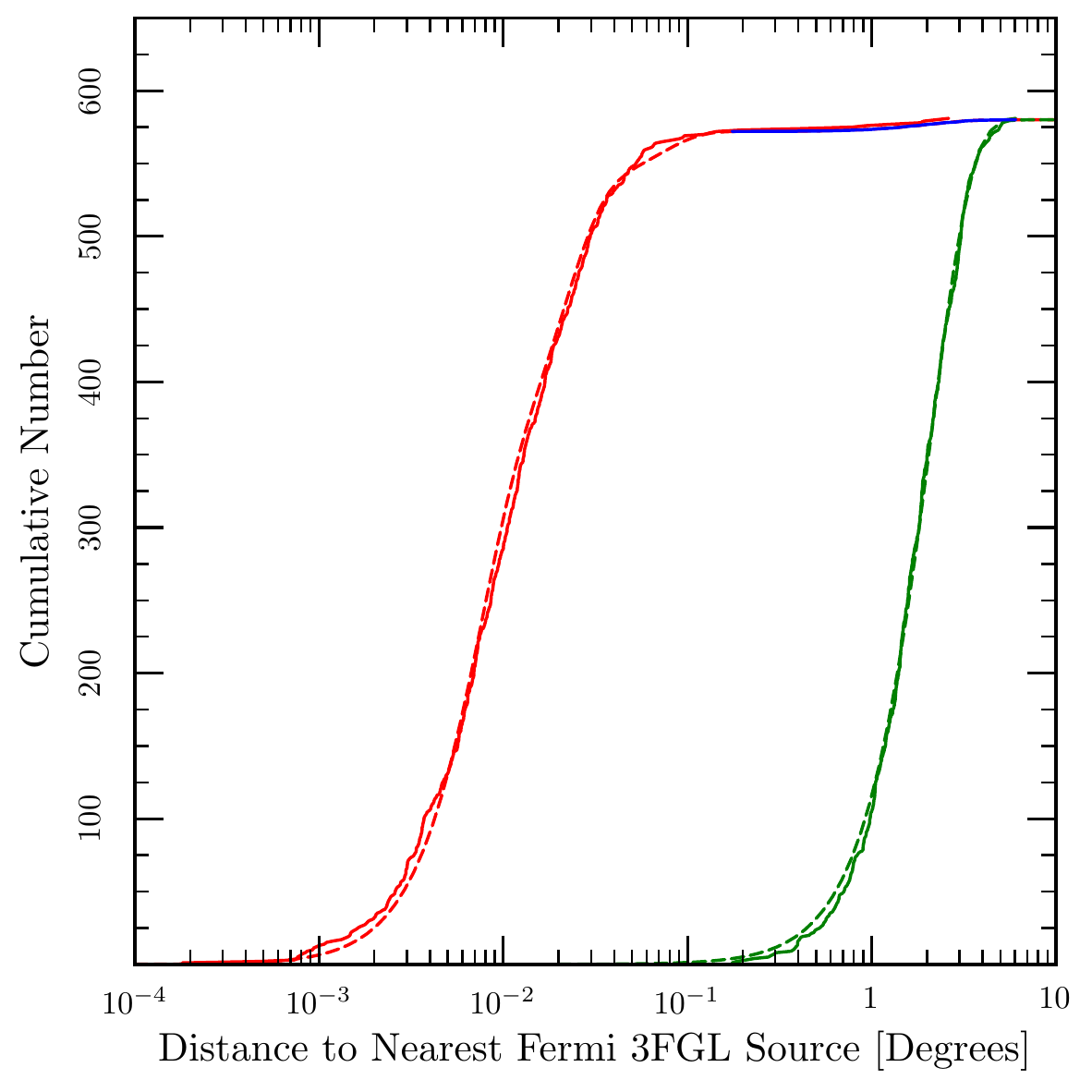}
\caption{The left panel gives the results for all sources, and the right
  panel has $|b|>10^\circ$. The distance from a Fermi FAST source to
  the nearest Fermi 3FGL source.  This demonstrates that ninety
  percent of the Fermi FAST sources are associated with sources in the
  Fermi 3FGL (across the whole sky) and nearly all at Galactic
  latitudes greater than ten degrees. The solid red curves give the
  observed cumulative distribution of nearest distances.  The solid green
  curves give the cumulative distribution that one would expect if
  there were no associated between the Fermi FAST and Fermi 3FGL
  sources.  This is calculated by performing the same analysis as the
  red curves but with the Galactic coordinates inverted.  The blue
  curve yields the false positive rate.  The dashed red and dashed green curves
  are Rayleigh distributions that are fit to the observed
  distributions.  The typical positional error between associated 3FGL
  and FAST sources is 1-2~arcminutes.  }
\label{fig:corresponances}
\end{figure*}

All of the sources detected by FermiFAST above five-sigma in the 1~GeV
and the 100~MeV sample are depicted in Fig.~\ref{fig:sources}.  One
can see that nearly all of the FermiFAST sources correspond to 3FGL
sources and in fact a few may correspond to several sources.  On the
other hand, there are many 3FGL sources that do not have counterparts
in the FermiFAST catalogue.  We can examine the statistics of the
correspondances more carefully by examining the distance across the
sky between the nearest neighbours in each of the two catalogues:
FermiFAST and 3FGL.  These are plotted as the red curves in
Fig.~\ref{fig:corresponances} and the dashed red curve is a multi-Rayleigh
distribution fit to the red curves.  We assume a Rayleigh distribution
to account for normally distributed errors in both directions along
the sky.  We find that the nearest neighbour in the 3FGL of almost all sources
in the FermiFAST catalogue lies within about 1-2
arcminutes.  We must assess whether these are chance coincidences.  If
the sources in the two catalogues are not correlated with each other
({\em i.e.} there are no real counterparts), then one would expect
that the cumulative distribution of nearest neighbour distances to
grow as a $1-\exp(\lambda \Omega)$ where $\lambda$ is the density of
sources on the sky and $\Omega$ is solid angle enclosed by a circle
centred on the object and passing through the nearest neighbour.
Given that there are 3,029 Fermi 3FGL sources on the sky, the
cumulative distribution in this case would approximately be
$1-\exp(-r^2/2\sigma^2)$ where $\sigma\approx 1.5^\circ$, so it is
unlikely that these counterparts at a typical distance of 1 arcminutes
are by chance.

We also can determine the distribution of unassociated pairs from the
data itself by inverting the coordinates of the sources in one
catalogue and looking for the nearest neighbours again.  To be precise
we change the sign of the Galactic latitude and longitude of each
source in the 3FGL and repeat the nearest neighbour search.  We invert
the coordinates instead of assigning random positions in order to
preserve the clustering of the sources on the sky and the density in
Galactic latitude and longitude.  Here all the correspondances are by
chance.  These results are given by the green curves in the various
panels and let us assess that nearly all the sources in the FermiFAST
have counterparts in the 3FGL.  In the all-sky map (left panel, we can
assess the number of FermiFAST sources that are not in the 3FGL by
fitting the cumulative distribution with several Rayleigh
distributions that quantify the positional error for the true
counterparts and the chance of a false counterpart determined by
fitting the cumulative distribution of false counterparts with a
Rayleigh distribution as well.  The distribution of false counterparts
(green curves in Fig.~\ref{fig:corresponances}) is well characterised
by a Rayleigh distribution with $\sigma=1.5^\circ$ (magenta curves) as
expected from the distribution of Fermi sources on the sky. We find
that 35 of the 800 sources do not have a counterpart in 3FGL.  If we
now focus on the right panel of Fig.~\ref{fig:corresponances} we find
that out of the 580 high-latitude sources in FermiFAST only eight lack
a counterpart in 3FGL.

\begin{figure}
\includegraphics[width=\columnwidth]{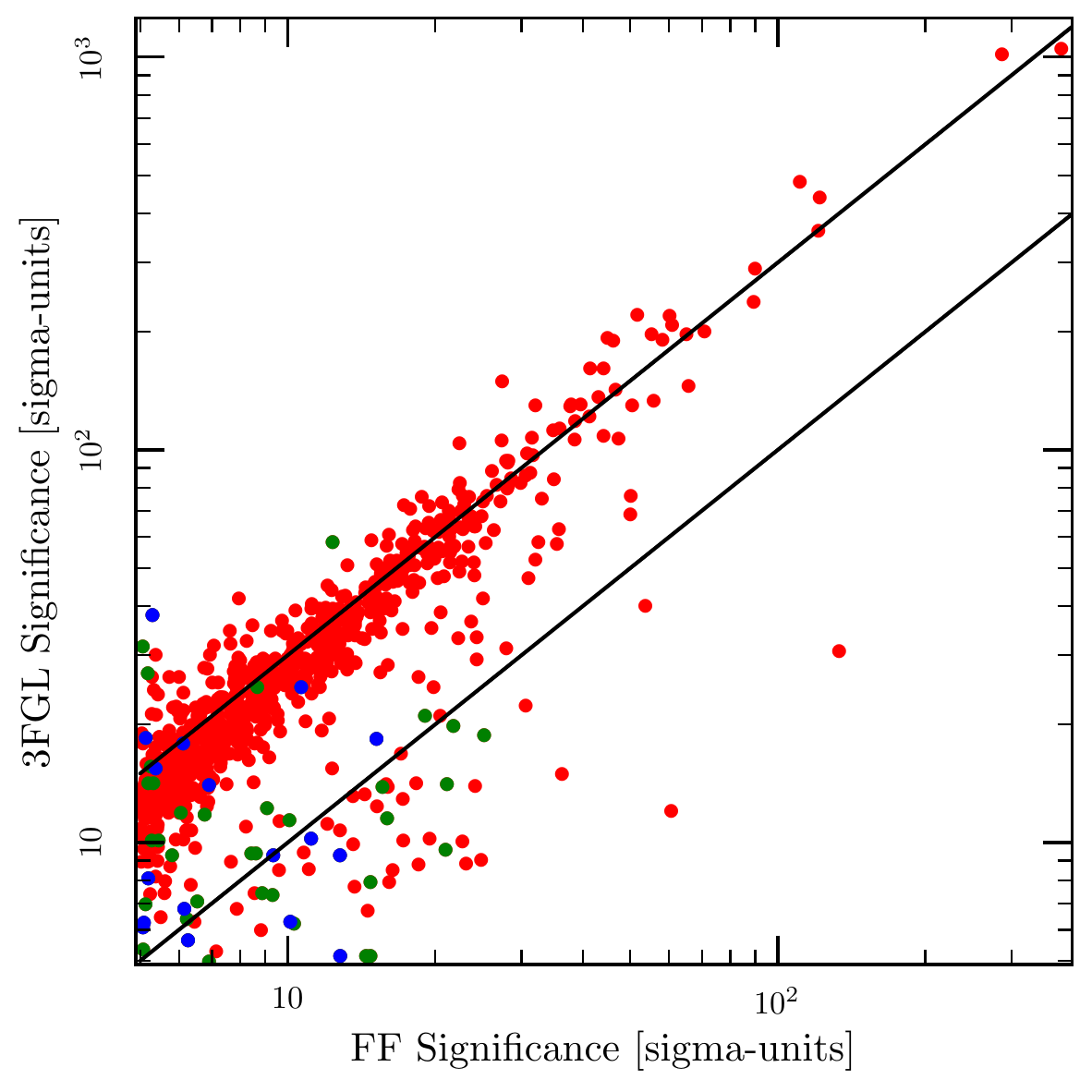}
\caption{3FGL Significances vs FermiFAST.  The red points are where
  the counterparts lie within $0.2^\circ$ of each other, the green the
  counterparts lies between $0.2^\circ$ and $1^\circ$, and the blue
  have counterparts further than $1^\circ$ away. The upper line traces
  a 3FGL signifance three times larger than the FermiFAST
  significance.  The lower line traces equal significance for the two techniques.
}
\label{fig:fermi_comp}
\end{figure}

One thing that is glaringly obvious from Fig.~\ref{fig:sources} is
that most sources in the 3FGL lack a counterpart in the FermiFAST
catalogue, in both the 1~GeV and the 100~MeV versions.  Given that the
algorithm outlined here is much less comprehensive that the techniques
used to generate the Fermi catalogues (there is no modelling of the
background, multiple sources or spectra), this is not surprising.
However, it is useful to figure out what sources are missing from the
FermiFAST catalogue and why.  Fig.~\ref{fig:fermi_comp} shows the
relationship between the significances of a source assigned by
FermiFAST and by the 3FGL.  In particular the 3FGL significances are
three times larger (the upper line) than FermiFAST typically, so if
both apply a threshold of five-sigma, FermiFAST will find fewer
sources.  However, there is a large population of sources for which
the two likelihoods are similar or the FermiFAST likelihood is larger.
Since FermiFAST does not perform any spectral modelling, perhaps these
sources have poor spectral fits in the 3FGL, yielding smaller
likelihoods.  Furthermore, this indicates a discovery space for
FermiFAST to find point sources where we do not have a good prior
notion of the spectral model.  The rightmost outlying point in the
lower-right is the Crab pulsar whose 3FGL likelihood is artificially
too low due to the way it is fit within the catalogue.  The second
rightmost point is 3FGL~J1745.6-2859c. It is likely that FermiFAST has
combined the significance of this 3FGL source with the nearby source
3FGL~J1745.3-2903c.  The latter source has a 3FGL significance of 20.6.

\begin{figure}
\includegraphics[width=\columnwidth]{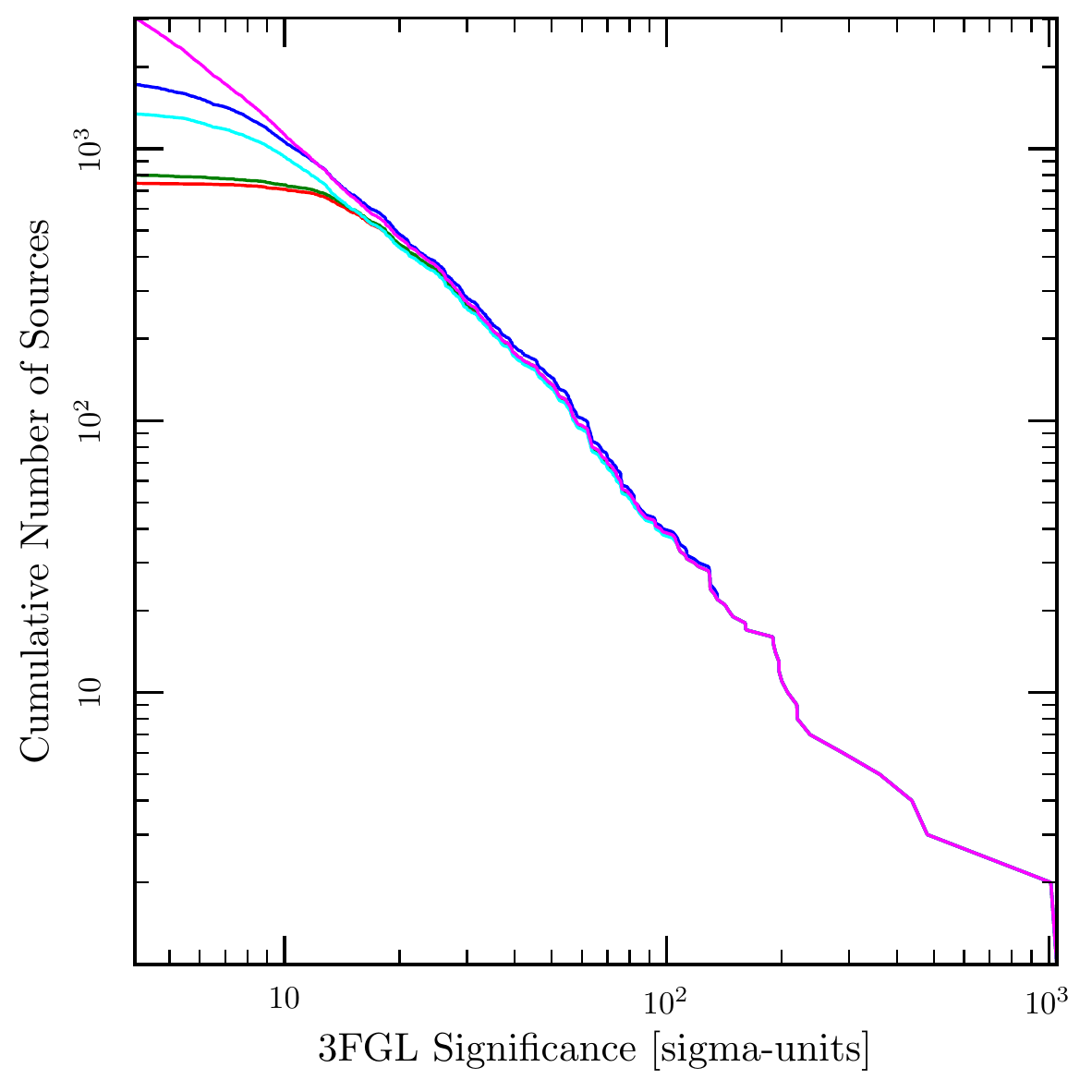}
\caption{
  Cumulative distribution of 3FGL significances for the entire 3FGL
  (blue) and the entire FermiFAST 1-GeV catalogue with a five-sigma
  cut (green) and FermiFAST where the 3FGL counterpart lies within
  $0.2^\circ$ (red) again with a five-sigma cut.  The blue and cyan
  give the same as green and red but with a three-sigma cut.}
\label{fig:sign-dist}
\end{figure}

To understand further whether FermiFAST is simply missing less
significant 3FGL sources we can look at the cumulative distribution of
3FGL significances of sources that appear in the FermiFAST 1~GeV
catalogue and all the 3FGL sources in Fig.~\ref{fig:sign-dist}.  We
see that FermiFAST catalogue is essentially complete for all 3FGL
sources above about 18-sigma, so FermiFAST can quickly (in six minutes)
generate a sample from the Fermi data stream of the upper quartile of
sources that would appear a Fermi catalogue using the full likelihood
technique to construct the catalogue.

\begin{figure}
  \includegraphics[width=\columnwidth]{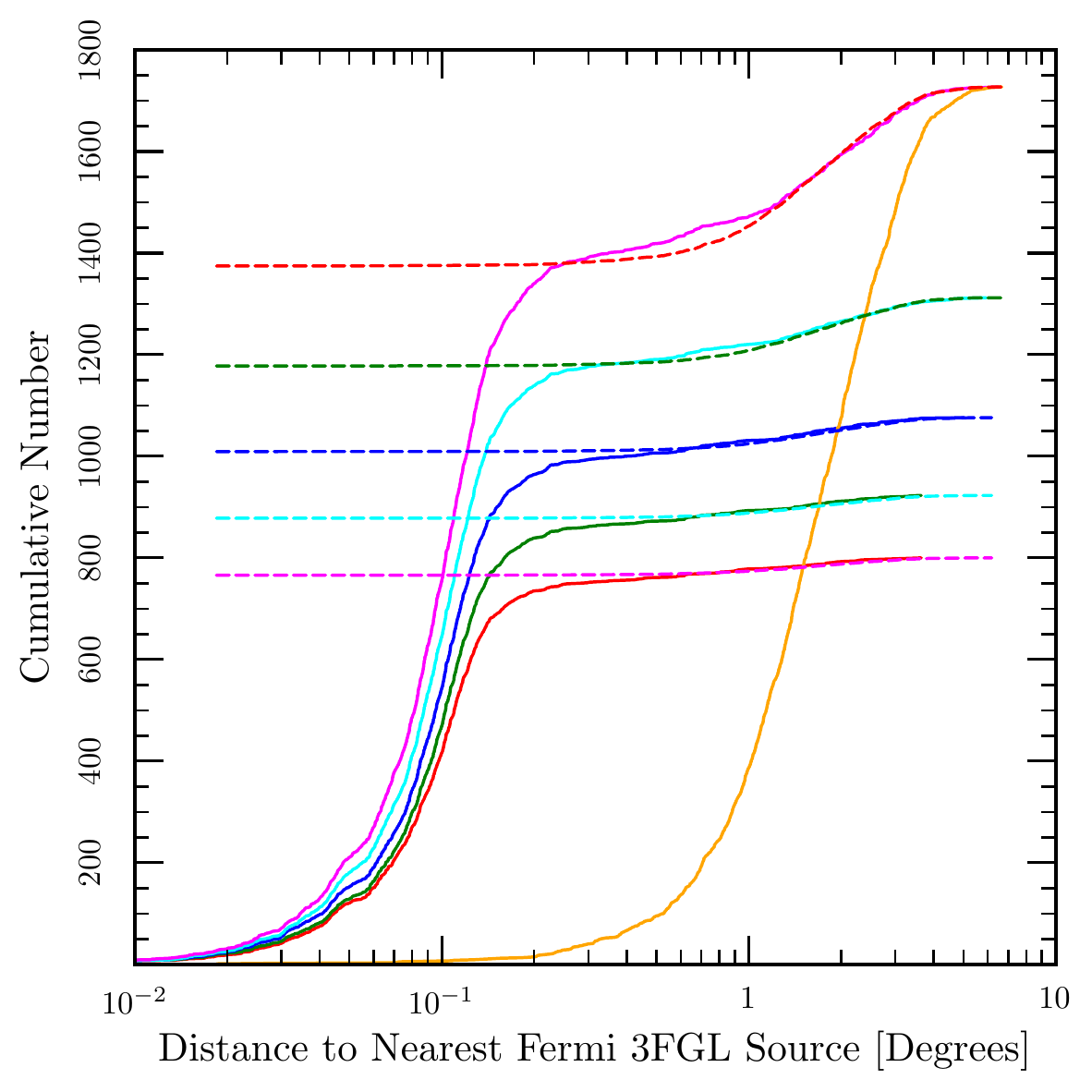}
  \caption{The cumulative distribution of distances between sources in
    FermiFAST and 3FGL for various significance thresholds from top
    to bottom of 3-sigma, 3.5-sigma, 4-sigma, 4.5-sigma and 5-sigma.
    The dashed curves that start above zero and join each cumulative
    distribution at large radii trace the cumulative distribution of
    distances for the false matches in each sample.}
  \label{fig:ffl_comp}
\end{figure}

The question arises: is it possible to do better? We can reduce the
significance threshold for find sources in the 1-GeV FermiFAST
catalogue.  We expect that the number of sources that do not have firm
associations in the 3FGL to increase but also for the completeness to
increase as well.  We can use the empirical distance distribution for
the false matches depicted in yellow in Fig.~\ref{fig:ffl_comp} to
split the distribution of the matches between FermiFAST and the 3FGL
statistcally into the true matches and the false matches to measure
the completeness and purity of the samples.  To do this we assume that
all the true matches are closer than any of the false matches and
use a ranked-sum-test to determine the fraction of true matches
from the observed distribution of matches (see Appendix for details).  For
the 1-GeV catalogue this appears to be a good assumption.  The
contribution of the false matches to the cumulative distributions is
depicted in the same colour as the measured distributions but dashed.
The distribution of the close matches here is somewhat different from that
in Fig.~\ref{fig:corresponances} because we have not used the improved
localizations to find the matches.  The localization improvement tends
to fail more often for sources of low significance (see
Fig.~\ref{fig:position}).  Tab.~\ref{tab:comppure} gives the results
of this trial.  The key result is that the five-sigma sample is very
pure; only about five percent of the sources lack firm associations in
the 3FGL.  On the other hand, if one is willing to sacrifice the
purity of the sample, one can achieve nearly 50\% completeness
relative to the 3FGL by reducing the significance threshold to
$3-\sigma$.  Fig.~\ref{fig:sign-dist} shows by accepting a lower
purity one can get a complete 3FGL sample to about ten-sigma,
comprising nearly half of the 3FGL sample.
\begin{table}
  \caption{Completeness and Purity}
  \label{tab:comppure}
  \begin{tabular}{l|rrrrr}
    \hline
    Threshold & \multicolumn{1}{c}{Sources} & \multicolumn{1}{c}{True} & \multicolumn{1}{c}{False} & \multicolumn{1}{c}{Purity} & \multicolumn{1}{c}{Comp} \\
    \hline
    $3-\sigma$   & 1727 & 1374 & 352 & 79.8\% & 48.4\% \\
    $3.5-\sigma$ & 1312 & 1178 & 134 & 89.8\% & 41.8\% \\
    $4-\sigma$   & 1076 & 1009 &  67 & 93.8\% & 36.2\% \\
    $4.5-\sigma$ &  923 &  877 &  46 & 95.1\% & 31.3\% \\
    $5-\sigma$   &  800 &  765 &  35 & 95.8\% & 27.0\%  
  \end{tabular}
\end{table}

We saw in Fig.~\ref{fig:fermi_comp} that in general the FermiFAST
significances and the 3FGL significances are well correlated.
However, for about one fifth of the sources $S(\mathrm{3FGL}) < 2
S(\mathrm{FF})$ and for about one tenth $S(\mathrm{3FGL}) <
S(\mathrm{FF})$.  This demonstrates that at least for a fraction of
the 3FGL the FermiFAST technique is more sensitive, so the sources
that do lie in the FermiFAST catalogue but not the 3FGL may just be a
class of sources to which the 3FGL is less sensitive, perhaps sources
for which the suite of spectral models employed in the construction of
the 3FGL do not apply.  On the other hand, these unassociated sources
could simply be regions where the diffuse background is especially
bright and clumpy.  Because the FermiFAST technique assumes that the
background is smooth on the scale of the PSF, it would find a source
where the background is locally strong.  We possibly see the opposite
of this effect in Fig.~\ref{fig:acorr} where FermiFAST finds negative
sources that are most likely places where the background is locally
weak.

\begin{figure}
  \includegraphics[width=\columnwidth]{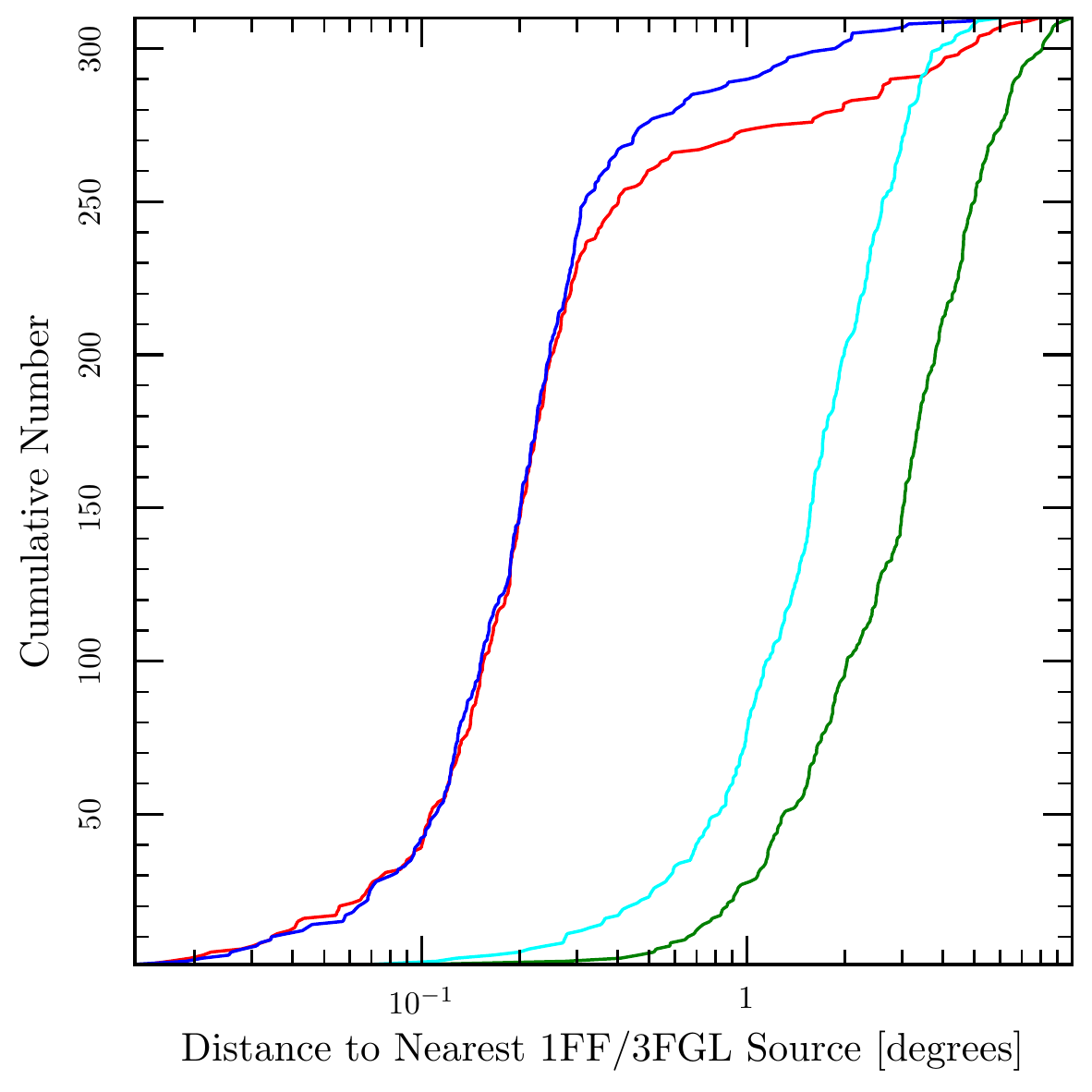}
  \caption{The cumulative distribution distances between sources in
    the 100-MeV catalogue and the 3FGL (blue, leftmost and higher) and
    the 1-GeV catalogue (red, second from left) and for the false
    matches between the 100-MeV and 3FGL (cyan, second from right) and
    100-MeV and 1-GeV catalogue (green, rightmost).}
  \label{fig:cum0.1GeV}
\end{figure}

It is apparent from Fig.~\ref{fig:sources} that there are fewer
100-MeV sources and that most of these sources coincide with 3FGL and
1-GeV FermiFAST sources.  In Fig.~\ref{fig:cum0.1GeV} the localization
is slightly poorer than Fig.~\ref{fig:ffl_comp} because we have used a
coarser HEALPix grid to sample the likelihoods. Moreover, for the 100-MeV
catalogue we have not optimized the likelihood to determine more
accurate positions because the optimization often drifts onto other
nearby detections for the more poorly localised low-energy photons.
Consequently, Fig.~\ref{fig:ffl_comp} is the best comparison
not Fig.~\ref{fig:corresponances}.  Again we have used a rank-sum test
to determine the sources that are likely to be shared between the
catalogues and also to determine those sources that appear in the
100-MeV catalogue but not the others.  The 100-MeV catalogue contains
312 sources of which 23 are in the 3FGL but not the 1-GeV catalogue,
16 are in the 1-GeV catagloue but not the 3FGL and 24 are in neither
of the other catalogues (249 are in all three). Relative to the 3FGL catalogue
the purity is 88.1\% and the completeness is 12.1\%. If one takes the
union of the five-sigma 1-GeV and the five-sigma 100-MeV FermiFAST
catalogues, there is a total of 788 sources that correspond to 836
3FGL sources ({\em i.e.} one FermiFAST source can correspond to
several 3FGL sources) for a slightly higher completeness fraction of
27.6\% than that of the five-sigma 1-GeV FermiFAST catalogue.

A final test that we perform is to examine the high-significance
FermiFAST detections that do not lie near sources within the 3FGL.
There are thirty-five of these detections.  The majority of these
detections (27) lie within 10~degrees of the Galactic plane.  Among
the entire five-sigma FermiFAST catalogue only 28\% lie within
10~degrees of the Galactic plane, so if the unassociated detections
within the Galactic plane were distributed as those associated with
3FGL sources, one would one expect 9.8 sources.  The probability of
finding at least 27 sources when one expects just 9.8 is $5\times
10^{-6}$.  In the entire 3FGL a similar fraction (27\%) of the sources
lie within 10~degrees of the Galactic plane, so again we find it
unlikely that the FermiFAST detections within the Galactic plane that
are not associated with 3FGL sources are distributed as the 3FGL
sources.  These probabilities and the fact that the gamma-ray
background is complicated near the Galactic plane point to the
conclusion that the unassociated detections near the Galactic plane are
regions of complicated background rather than true point sources.

To understand the unassociated detections outside of the Galactic
plane we analyse the Fermi data for the same four-year span, weeks 9
through 216, from 2008 August 4 to 2012 July 26. There are eight
unassociated detections with $|b|>10^{\circ}$ out of a total of 580
FermiFAST detections outside of the Galactic plane.  We include
photons within a region of interest of 20 degrees about the FermiFAST
detection and include 3FGL sources out to 30 degrees in the likelihood
fit. We generate the input model with \texttt{make3FGLxml.py}.  All
3FGL sources with significances greater than four are included in the
fit.  Those sources within 20 degrees of the detection and with
significances greater than five are allowed to vary in the fit.
Otherwise, the values found in the 3FGL catalogue are used.  We
perform a binned likelihood analysis for photons with energies between
100~MeV and 300~GeV.  We used the Pass-8 data with the isotropic
background model \texttt{iso\_P8R2\_SOURCE\_V6\_v06} and the Galactic
background model \texttt{gll\_iem\_v06}.  We choose the Pass-8 data,
so that we could use the latest background and extended emission
models to get the most accurate characterization of the FermiFAST
detections.  The results for Pass-7 data are similar. We allow the
normalization of the background models to vary but not the slope.
Tab.~\ref{tab:fermisig100MeV} presents the results for the binned
likelihhod fit for all the unassociated FermiFAST detections outside
of the Galactic plane.  The range of spectral indices is typical for
3FGL sources.  Two detections of particular interest are 5 and 24.
Detection 5 resulted in a negative TS value.  It is within the Large
Magellanic Cloud, so FermiFAST possibly identified a knot of gamma-ray
emission from within the galaxy.  Detection 24 also yielded a low
significance.  It lies within 30 arcminutes of the relatively weak
source 3FGL J0546.4+0031c.  This source has a significance of 5.0 in
the 3FGL, so although it was included in the fit, its parameters were
not allowed to vary.  The remaining detections all lie at least 50
arcminutes away from 3FGL sources and appear to be new sources that
were not found in the 3FGL catalogue. Each has a spectral index
between $2.0$ and $2.6$, near the mode of the distribution within the
3FGL catalogue, so they do not appear unusual in that context.  Of the
eight unassociated FermiFAST detections, six have significances
greater than five when a detailed binned likelihood analysis is
performed, yielding a final contamination rate of 0.3\%.

\begin{table*}
  \caption{ All of the FermiFAST detections that lack a 3FGL
    counterpart and lie outside the Galactic plane ($|b|>10^{\circ}$),
    results from a four-year binned likelihood analysis for photons
    between 100~MeV and 300~GeV.  The FermiFAST columns have the same
    meanings as in Tab.~\ref{tab:topten}. The $S(100)$ column gives
    the significance of the binned likelihood fit in sigma-units.  The
    $\Gamma$ column gives the spectral index, $dN/dE \propto E^{-\Gamma}$,
    and the flux is the total photon flux from 100~MeV to 300~GeV in units
  of $10^{-7}\mathrm{photons}~\mathrm{cm}^{-2}~\mathrm{s}^{-1}$.}
  \label{tab:fermisig100MeV}
\begin{tabular}{rrrrrrrrrrrrrr}
    \hline
    \multicolumn{1}{c}{} & \multicolumn{1}{c}{RA} & \multicolumn{1}{c}{Dec}  & \multicolumn{1}{c}{$N_\gamma$}  &\multicolumn{1}{c}{$\bar r^2$} & \multicolumn{1}{c}{$\bar f$} & \multicolumn{1}{c}{$S(r^2)$} & \multicolumn{1}{c}{$S(f)$} & \multicolumn{1}{c}{$A_f$}  & \multicolumn{1}{c}{$A_\mathrm{PSF}$} & \multicolumn{1}{c}{$S(\mathrm{FF})$} & \multicolumn{1}{c}{$S(100)$}  & \multicolumn{1}{c}{$\Gamma$}  & \multicolumn{1}{c}{Flux}
 \\
 \hline
 5 & $ 84.56$ & $-69.20$ &   1960 & $0.381$ & $0.733$ & $ -18.3$ & $  35.7$ & $0.34$  & $0.32$ & $    15.18$ & \multicolumn{3}{c}{Not significant}  \\
10 & $232.90$ & $-56.21$ &  17522 & $0.461$ & $0.814$ & $ -17.9$ & $ 144.0$ & $0.11$  & $0.08$ & $    12.78$ & $10.2$ & $2.01\pm0.07$ & $16.3\pm2.6$\\
18 & $ 85.43$ & $ -1.94$ &   2775 & $0.433$ & $0.790$ & $ -12.3$ & $  52.9$ & $0.19$  & $0.15$ & $     9.31$ & $15.1$ & $2.47\pm0.09$ & $10.3\pm1.2$\\
20 & $ 85.39$ & $ -8.22$ &   3419 & $0.434$ & $0.793$ & $ -13.4$ & $  59.3$ & $0.18$  & $0.13$ & $     8.87$ & $14.0$ & $2.52\pm0.01$ & $10.6\pm0.2$\\ 
24 & $ 86.84$ & $  0.16$ &   2659 & $0.457$ & $0.805$ & $  -7.8$ & $  54.5$ & $0.12$  & $0.11$ & $     6.91$ & $2.89$  & $2.07\pm0.06$ & $1.2\pm0.2$\\
29 & $ 52.28$ & $ 31.20$ &   2493 & $0.453$ & $0.802$ & $  -8.2$ & $  52.2$ & $0.14$  & $0.10$ & $     6.12$ & $6.57$ & $2.18\pm0.27$ & $2.6\pm1.3$ \\ 
31 & $ 70.17$ & $ 25.63$ &   3275 & $0.471$ & $0.808$ & $  -5.7$ & $  61.0$ & $0.09$  & $0.08$ & $     5.93$ & $15.1$ & $2.42\pm0.09$ & $11.9\pm1.5$ \\
43 & $ 64.51$ & $ 28.29$ &   2260 & $0.457$ & $0.805$ & $  -7.1$ & $  50.3$ & $0.12$  & $0.09$ & $     5.09$ & $11.7$  & $2.40\pm0.18$ & $7.2\pm1.7$\\ 
\end{tabular}
\end{table*}

We looked for the nearest blazars to each of the unassociated
FermiFAST detections (Tab.~\ref{tab:fermisig100MeV}) in the Roma-BZCAT
catalogue of blazars \citep{2009A&A...495..691M}.  The results are
listed in Tab.~\ref{tab:roma}.  The closest objects in the Roma-BZCAT
to each of the unassociated FermiFAST detections are about one degree.
Given the density of the blazars in this catalogue on the sky, this is
the typical distance for the nearest source to be by chance;
consequently, we find that it is unlikely that any of FermiFAST
detections that did not have counterparts in the 3FGL have
counterparts in the Roma-BZCAT.

\begin{table}
  \caption{The nearest blazars found in the Roma-BZCAT \citep{2009A&A...495..691M} to
    the unassociated FermiFAST sources with the relative distance in degrees.}
  \label{tab:roma}
\centering
\begin{tabular}{rrrrl}
    \hline
    \multicolumn{1}{c}{} &     \multicolumn{1}{c}{Distance} & \multicolumn{1}{c}{RA} & \multicolumn{1}{c}{Dec} & \multicolumn{1}{c}{Roma-BZCAT} \\
        \hline
 5   & 0.70 & $77.1762$ & $+84.5346$    & BZBJ0508$+$8432 \\
10   & 1.65 & $50.5329$ & $-52.0933$    & BZQJ0322$-$5205 \\
18   & 1.15 & $77.1762$ & $+84.5346$    & BZBJ0508$+$8432 \\
20   & 1.12 & $77.1762$ &  $+84.5346$   & BZBJ0508$+$8432 \\
24   & 1.56 & $105.6367$ & $+85.8312$   & BZQJ0702$+$8549 \\
29   & 5.58 & $43.4900$  & $+51.0490$   & BZQJ0253$+$5102 \\
31   & 3.53 & $76.9842$  & $+67.6234$   & BZBJ0507$+$6737 \\
43   & 3.57 & $72.3471$  & $+63.5359$   & BZQJ0449$+$6332 \\
\end{tabular}
\end{table}

\section{Discussion}
\label{sec:discussion}

We have outlined a technique that can rapidly find point sources from
the photon data from the Fermi telescope.  If one focusses on the
photons above 1~GeV, one can reproduce about half of the 3FGL using
the 3FGL data set in about six minutes with a false association rate
of less than twenty percent. If one focuses on detections with
significance greater than five sigma, the contamination rate outside
of the Galactic plane is a few parts per thousand. There are several
immediate applications of this technique that come to mind as well as
several avenues of potential improvements to increase the sensitivity
and specificity and the data that we learn about each source.  First,
we will discuss some potential applications.

The speed of the technique opens several possible exploratory research
paths.  One can search fractions of the data stream for potential
sources that were active for a portion of the complete 3FGL observation
time and so did not reach the threshold for detectability over the
entire observation but were sufficiently bright to be detected over a
shorter window.  Given the large variability of blazars which form the
bulk of the sources outside the plane of the Galaxy, this is an
interesting avenue.  In fact although the 3FGL catalogue contains many
more sources than the 2FGL, many sources are absent in the 3FGL that
were in the 2FGL.  Some of these may have been spurious, but others
could have simply been variable, so the search for variable sources
provides an exciting avenue.

A second important direction is to use the FermiFAST algorithm to
generate a catalogue of sources, most likely not as deep as the 3FGL
and then use the catalogue for further statistical analysis of the
various populations.  Now what are the advantages of this when one
considers that the 3FGL already exists?  Because the FermiFAST
catalogue only requires a few minutes to generate one can explore the
statistical biases, selection and uncertainities in the catalogue
through simulated data much more easily than for the full 3FGL
catalogue.  A first step would be to insert artificial sources of
known fluxes and spectra into the data stream.  This could be down in
parallel by again inserting the sources on a HEALPix grid so that the
regions of interest for each source (Fig.~\ref{fig:expmap}) do not
overlap over the energy range considered.  Now we can measure the
relationship between the flux and significance throughout the sky to
obtain the selection function for the survey.  This is crucial
ingredient to determine the luminosity function of the underlying
population.  Because the underlying sources may be variable, we could
also insert variable sources into the data stream as well and combine
this with the time slicing analysis described in the previous
paragraph and develop for example a catalogue of blazars as a function
of mean luminosity and variability that is both statistically well
characterized and possibly contains new objects that were not
discovered in the previous Fermi catalogues.  Of course, many of the sources
will have already been detected by Fermi and possibly followed up
with further measurements, the key element here would be that the sample
would be statistically well characterized, so clear conclusions about the
population could be obtained.

Using the existing data stream without inserting sources, one could
characterize whether the measurement significances of the individual
sources are accurate measurements of Poisson likelihood.  To be specific
we would estimate the distribution of the values $\log L$ throughout
the sky so that a given value of $\log L$ could be associated with
the probability of the null hypothesis of no source. The
first step would be to resample the photon data stream through a
bootstrap --- this would generate a new list of photons by selecting
the same number of photons from the original list but with
replacement, so that the same photon could appear many times in the
new sample and of course some photons won't appear at all.  Each new
list of photons would generate a new catalogue to determine what
properties of the sources and the catalogue as a whole are robust
against the Poisson fluctuations in the data stream.  This would
verify the underlying assumptions of the significances of the sources
as well as determine how sources near the detection threshold enter
and leave the sample.  Because the sample is dominated by the faintest
sources near the detection threshold (see Fig.~\ref{fig:fermi_comp}),
understanding this effect is again crucial to understand the sample.

To detect point sources from the Fermi photon data stream, we only
used the data stream and the estimates of the point-spread
function. There is additional data available that could help to
improve the sensitivity of the technique, reduce the number of false
positives and better characterize the sources detected.  First, the
likelihood function, Eq.~\ref{eq:14}, assumes that the gamma-ray
background is smooth everywhere.  However, we know from independent
measurements, for example the Galactic synchrotron emission, that the
gamma-ray background is likely to be clumpy and the Fermi team
provides estimates of this background \citep[e.g.][]{Fermi1602.07246}.
By combining this background
estimate along with the integral of the effective area and time that a
particular direction and energy was observed, we can develop an
improved likelihood function that includes the estimated background.
This could reduce both the number of candidate sources detected with
$A_\mathrm{PSF}<0$ and the number of candidate sources without
counterparts in the 3FGL.  The addition of this information would not
dramatically increase the time to construct the catalogue (we have
already included this functionality in the software but it is not yet
well characterized).

One could also take an alternative route.  Looking again at
Eq.~\ref{eq:14}, we can use the relative size of the two terms in the
brackets to assign a fraction of each photon to the source and to the
background.  Again if we include information on the effective area and
the exposure time, we obtain an estimate of the spectrum from each
source and from the background.  Because a given photon could be
assigned to several sources as well as the background, we would also
have to develop a good statistical model to split the photon among the
sources.  Of course, a first estimate could be obtained by ignoring
the fact that several potential sources could claim a given photon and
resolve the gamma-ray data into sources and background \citep[as done
 by][]{2015A&A...581A.126S}. This would provide an independent
confirmation of the structure of the gamma-ray background, but perhaps
more importantly it would yield an estimate of the spectrum of each
source. We have performed a fit to the source spectrum using the
standard Fermi analysis after the sources have been detected with
FermiFAST.  However, by including the fit in the FermiFAST analysis,
we could take advantage of the photon database to accelerate the
calculation.  Including the fit in the FermiFAST detection could
improve both the sensitivity as we could find fainter sources against
the background if they follow one of the spectral models and the
specificity by excluding potential sources that do not follow the
models.

\section{Conclusions}
\label{sec:conclusions}

FermiFAST provides the infrastructure both for exploratory analysis of
the Fermi data to find new sources and new variable sources and also
to create a sample of sources for population studies that can be
characterized comprehensively by Monte Carlo simulation of the
detection technique.  The latter may allow new understanding of the
population of Galactic and extragalactic gamma-ray sources and their
evolution. In this paper we have purposefully used the Pass 7 response
functions and just four years of data to compare with the 3FGL.
Nearly an additional four years of data are now available with the
Pass 8 reductions \citep{2013arXiv1303.3514A}, so a sneak preview of
the 4FGL would be a very natural next step.

\section*{Acknowledgments}

Jeremy Heyl would like to thank Elisa Antolini for the conversations
that formed the impetus for this paper.  The software used in this
paper and the catalogues generated are available at
\url{http://ubc-astrophysics.github.io}.  We used the VizieR Service,
the NASA ADS service, the Fermi Science Support Center, the
astrometry.net $k-d$~tree library, the HEALPix and HEALPy libraries
and arXiv.org. This work was supported by the Natural Sciences and
Engineering Research Council of Canada, the Canadian Foundation for
Innovation, the British Columbia Knowledge Development Fund and the
Bertha and Louis Weinstein Research Fund at the University of British
Columbia.

\bibliography{fermi}

\begin{thebibliography}{}

\bibitem[\protect\citeauthoryear{{Abdo}, {Ackermann}, {Ajello}, {Allafort},
  {Antolini}, {Atwood}, {Axelsson}, {Baldini}, {Ballet}, {Barbiellini} \& et
  al.}{{Abdo} et~al.}{2010}]{2010ApJS..188..405A}
{Abdo} A.~A.,  {Ackermann} M.,  {Ajello} M.,  {Allafort} A.,  {Antolini} E.,
  {Atwood} W.~B.,  {Axelsson} M.,  {Baldini} L.,  {Ballet} J.,  {Barbiellini}
  G.,    et al. 2010, \apjs, 188, 405

\bibitem[\protect\citeauthoryear{{Abdo}, {Ajello}, {Allafort}, {Baldini},
  {Ballet}, {Barbiellini}, {Baring}, {Bastieri}, {Belfiore}, {Bellazzini} \& et
  al.}{{Abdo} et~al.}{2013}]{2013ApJS..208...17A}
{Abdo} A.~A.,  {Ajello} M.,  {Allafort} A.,  {Baldini} L.,  {Ballet} J.,
  {Barbiellini} G.,  {Baring} M.~G.,  {Bastieri} D.,  {Belfiore} A.,
  {Bellazzini} R.,    et al. 2013, \apjs, 208, 17

\bibitem[\protect\citeauthoryear{{Acero}, {Ackermann}, {Ajello}, {Albert},
  {Atwood}, {Axelsson}, {Baldini}, {Ballet}, {Barbiellini} et~al.,}{{Acero}
  et~al.}{2015}]{2015ApJS..218...23A}
{Acero} F.,  {Ackermann} M.,  {Ajello} M.,  {Albert} A.,  {Atwood} W.~B.,
  {Axelsson} M.,  {Baldini} L.,  {Ballet} J.,  {Barbiellini} G.,    et~al.,
  2015, \apjs, 218, 23

\bibitem[\protect\citeauthoryear{Acero et~al.,}{Acero
  et~al.}{2016}]{Fermi1602.07246}
Acero F.,  et~al., 2016, \apjs, accepted (1602.07246)

\bibitem[\protect\citeauthoryear{{Ackermann}, {Ajello}, {Albert}, {Allafort},
  {Atwood}, {Axelsson}, {Baldini}, {Ballet}, {Barbiellini}, {Bastieri}
  et~al.,}{{Ackermann} et~al.}{2012}]{2012ApJS..203....4A}
{Ackermann} M.,  {Ajello} M.,  {Albert} A.,  {Allafort} A.,  {Atwood} W.~B.,
  {Axelsson} M.,  {Baldini} L.,  {Ballet} J.,  {Barbiellini} G.,  {Bastieri}
  D.,    et~al., 2012, \apjs, 203, 4

\bibitem[\protect\citeauthoryear{{Ackermann}, {Ajello}, {Allafort}, {Antolini},
  {Atwood}, {Axelsson}, {Baldini}, {Ballet}, {Barbiellini},
  et~al.,}{{Ackermann} et~al.}{2011}]{2011ApJ...743..171A}
{Ackermann} M.,  {Ajello} M.,  {Allafort} A.,  {Antolini} E.,  {Atwood} W.~B.,
  {Axelsson} M.,  {Baldini} L.,  {Ballet} J.,  {Barbiellini} G.,     et~al.,
  2011, \apj, 743, 171

\bibitem[\protect\citeauthoryear{{Ackermann}, {Ajello}, {Allafort}, {Asano},
  {Atwood}, {Baldini}, {Ballet}, {Barbiellini}, {Bastieri}, {Bechtol}
  et~al.,}{{Ackermann} et~al.}{2013}]{2013ApJ...765...54A}
{Ackermann} M.,  {Ajello} M.,  {Allafort} A.,  {Asano} K.,  {Atwood} W.~B.,
  {Baldini} L.,  {Ballet} J.,  {Barbiellini} G.,  {Bastieri} D.,  {Bechtol} K.,
     et~al., 2013, \apj, 765, 54

\bibitem[\protect\citeauthoryear{{Ackermann}, {Ajello}, {Allafort}, {Atwood},
  {Baldini}, {Ballet}, {Barbiellini}, {Bastieri}, {Bechtol}, {Belfiore}
  et~al.,}{{Ackermann} et~al.}{2013}]{2013ApJS..209...34A}
{Ackermann} M.,  {Ajello} M.,  {Allafort} A.,  {Atwood} W.~B.,  {Baldini} L.,
  {Ballet} J.,  {Barbiellini} G.,  {Bastieri} D.,  {Bechtol} K.,  {Belfiore}
  A.,    et~al., 2013, \apjs, 209, 34

\bibitem[\protect\citeauthoryear{{Atwood}, {Albert}, {Baldini}, {Tinivella},
  {Bregeon}, {Pesce-Rollins}, {Sgr{\`o}}, {Bruel}, {Charles} et~al.,}{{Atwood}
  et~al.}{2013}]{2013arXiv1303.3514A}
{Atwood} W.,  {Albert} A.,  {Baldini} L.,  {Tinivella} M.,  {Bregeon} J.,
  {Pesce-Rollins} M.,  {Sgr{\`o}} C.,  {Bruel} P.,  {Charles} E.,    et~al.,
  2013, ArXiv e-prints

\bibitem[\protect\citeauthoryear{Bentley}{Bentley}{1975}]{Bentley:1975:MBS:361002.361007}
Bentley J.~L.,  1975, Commun. ACM, 18, 509

\bibitem[\protect\citeauthoryear{{Braun}, {Dumm}, {De Palma}, {Finley}, {Karle}
  \& {Montaruli}}{{Braun} et~al.}{2008}]{2008APh....29..299B}
{Braun} J.,  {Dumm} J.,  {De Palma} F.,  {Finley} C.,  {Karle} A.,
  {Montaruli} T.,  2008, Astroparticle Physics, 29, 299

\bibitem[\protect\citeauthoryear{{Campana}, {Bernieri}, {Massaro}, {Tinebra} \&
  {Tosti}}{{Campana} et~al.}{2013}]{2013Ap&SS.347..169C}
{Campana} R.,  {Bernieri} E.,  {Massaro} E.,  {Tinebra} F.,    {Tosti} G.,
  2013, \apss, 347, 169

\bibitem[\protect\citeauthoryear{{Campana}, {Massaro}, {Gasparrini}, {Cutini}
  \& {Tramacere}}{{Campana} et~al.}{2008}]{2008MNRAS.383.1166C}
{Campana} R.,  {Massaro} E.,  {Gasparrini} D.,  {Cutini} S.,    {Tramacere} A.,
   2008, \mnras, 383, 1166

\bibitem[\protect\citeauthoryear{{Cash}}{{Cash}}{1979}]{1979ApJ...228..939C}
{Cash} W.,  1979, \apj, 228, 939

\bibitem[\protect\citeauthoryear{{Ciprini}, {Tosti}, {Marcucci}, {Cecchi},
  {Discepoli}, {Bonamente}, {Germani}, {Impiombato}, {Lubrano} \&
  {Pepe}}{{Ciprini} et~al.}{2007}]{2007AIPC..921..546C}
{Ciprini} S.,  {Tosti} G.,  {Marcucci} F.,  {Cecchi} C.,  {Discepoli} G.,
  {Bonamente} E.,  {Germani} S.,  {Impiombato} D.,  {Lubrano} P.,    {Pepe} M.,
   2007, in {Ritz} S.,  {Michelson} P.,   {Meegan} C.~A.,  eds, The First GLAST
  Symposium Vol.~921 of American Institute of Physics Conference Series, {1D,
  2D, 3D wavelet methods for gamma-ray source analysis}.
pp 546--547

\bibitem[\protect\citeauthoryear{{Damiani}, {Maggio}, {Micela} \&
  {Sciortino}}{{Damiani} et~al.}{1997a}]{1997ApJ...483..350D}
{Damiani} F.,  {Maggio} A.,  {Micela} G.,    {Sciortino} S.,  1997a, \apj, 483,
  350

\bibitem[\protect\citeauthoryear{{Damiani}, {Maggio}, {Micela} \&
  {Sciortino}}{{Damiani} et~al.}{1997b}]{1997ApJ...483..370D}
{Damiani} F.,  {Maggio} A.,  {Micela} G.,    {Sciortino} S.,  1997b, \apj, 483,
  370

\bibitem[\protect\citeauthoryear{{G{\'o}rski}, {Hivon}, {Banday}, {Wandelt},
  {Hansen}, {Reinecke} \& {Bartelmann}}{{G{\'o}rski}
  et~al.}{2005}]{2005ApJ...622..759G}
{G{\'o}rski} K.~M.,  {Hivon} E.,  {Banday} A.~J.,  {Wandelt} B.~D.,  {Hansen}
  F.~K.,  {Reinecke} M.,    {Bartelmann} M.,  2005, \apj, 622, 759

\bibitem[\protect\citeauthoryear{Lang}{Lang}{2009}]{LangPhD}
Lang D.,  2009, PhD thesis, University of Toronto

\bibitem[\protect\citeauthoryear{{Massaro}, {Giommi}, {Leto}, {Marchegiani},
  {Maselli}, {Perri}, {Piranomonte} \& {Sclavi}}{{Massaro}
  et~al.}{2009}]{2009A&A...495..691M}
{Massaro} E.,  {Giommi} P.,  {Leto} C.,  {Marchegiani} P.,  {Maselli} A.,
  {Perri} M.,  {Piranomonte} S.,    {Sclavi} S.,  2009, \aap, 495, 691

\bibitem[\protect\citeauthoryear{{Massaro}, {Tinebra}, {Campana} \&
  {Tosti}}{{Massaro} et~al.}{2009}]{2009arXiv0912.3843M}
{Massaro} E.,  {Tinebra} F.,  {Campana} R.,    {Tosti} G.,  2009, ArXiv
  e-prints

\bibitem[\protect\citeauthoryear{{Mattox}, {Bertsch}, {Chiang}, {Dingus},
  {Digel}, {Esposito}, {Fierro}, {Hartman}, {Hunter}, {Kanbach}, {Kniffen}
  et~al.,}{{Mattox} et~al.}{1996}]{1996ApJ...461..396M}
{Mattox} J.~R.,  {Bertsch} D.~L.,  {Chiang} J.,  {Dingus} B.~L.,  {Digel}
  S.~W.,  {Esposito} J.~A.,  {Fierro} J.~M.,  {Hartman} R.~C.,  {Hunter} S.~D.,
   {Kanbach} G.,  {Kniffen} D.~A.,    et~al., 1996, \apj, 461, 396

\bibitem[\protect\citeauthoryear{{Nolan}, {Abdo}, {Ackermann}, {Ajello},
  {Allafort}, {Antolini}, {Atwood}, {Axelsson}, {Baldini}, {Ballet} \& et
  al.}{{Nolan} et~al.}{2012}]{2012ApJS..199...31N}
{Nolan} P.~L.,  {Abdo} A.~A.,  {Ackermann} M.,  {Ajello} M.,  {Allafort} A.,
  {Antolini} E.,  {Atwood} W.~B.,  {Axelsson} M.,  {Baldini} L.,  {Ballet} J.,
    et al. 2012, \apjs, 199, 31

\bibitem[\protect\citeauthoryear{{Schmidt}}{{Schmidt}}{1968}]{1968ApJ...151..393S}
{Schmidt} M.,  1968, \apj, 151, 393

\bibitem[\protect\citeauthoryear{{Selig}, {Vacca}, {Oppermann} \&
  {En{\ss}lin}}{{Selig} et~al.}{2015}]{2015A&A...581A.126S}
{Selig} M.,  {Vacca} V.,  {Oppermann} N.,    {En{\ss}lin} T.~A.,  2015, \aap,
  581, A126

\bibitem[\protect\citeauthoryear{{Sutherland} \& {Saunders}}{{Sutherland} \&
  {Saunders}}{1992}]{1992MNRAS.259..413S}
{Sutherland} W.,  {Saunders} W.,  1992, \mnras, 259, 413

\end{thebibliography}
\bibliographystyle{mn2e}

\section*{Appendix}

We use a ranked-sum test to estimate the fraction of coincidences
between members of two catalogues that are likely to be true
counterparts rather that chance coincidences.  We have the
distribution of the nearest-neighbour separations between the two
catalogues and also an estimate of the distribution of the
nearest-neighbour separations for the chance coincidences.  We
furthermore assume that the true matches all have smaller separations
than the chance coincidences.

The basic idea is to compare the list of observed separations to the
list of separations of chance coincidences in pair-by-pair match ups
where the winner is the member of the pair that is smaller.  If there
were actually no true counterparts in the observed list, then the
observed list and the list of chance coincidences would have the same
distribution and one would expect the half of pairwise matches to be
won by members of the observed list and half to be won by members of
the chance list.  On the other hand, if the entire observed list
consists of true matches, then the members of the observed list should
win all of the time, so we have
$$
f_\textrm{\scriptsize True Matches} = 2 f_\textrm{\scriptsize Observed Wins} - 1.
$$
This formula holds even if the two lists have differing numbers of
elements.

To calculate the fraction of wins, we do not actually perform a pair-by-pair
comparision which would require ${\cal O}(N^2)$ comparisons but rather sort the
two lists together, ${\cal O}(N \ln N)$, and keep track of which elements
come from which list as in the following Python snippet:
\begin{verbatim}
rall=np.concatenate((r1,r2))
n1all=np.concatenate((np.ones(len(r1)),
                      np.zeros(len(r2))))
n2all=np.concatenate((np.zeros(len(r1)),
                      np.ones(len(r2))))
index=np.argsort(rall)
n1all=n1all[index]
n2all=n2all[index]
c1all=np.cumsum(n1all)
obswins=np.sum(n2all*c1all)
frac=2*obswins/(len(r1)*len(r2))-1
\end{verbatim}
where ${\tt r1}$ contains the list of observed separations of the
nearest neighbours in the two catalogues and ${\tt r2}$ contains the
list of separations of the chance nearest neighbours.
    
\label{lastpage}

\end{document}